\documentclass[oneside,12pt]{{article}}
\usepackage[a4paper,margin=1in]{geometry}
\usepackage{tikz}
\usepackage{lipsum}  
\usepackage{lineno}
\usepackage{fmtcount} 
\usepackage{array,multirow}
\usepackage{longtable}
\usepackage{afterpage}
\usepackage{lscape}
\usepackage[export]{adjustbox}
\modulolinenumbers[5]
\usepackage{listings}
\usepackage{xcolor}
\usepackage[authoryear]{natbib}
\usepackage{mathtools}
\usepackage{appendix}
\usepackage{verbatim}
\usepackage{amsmath,bm}
\usepackage{authblk}
\usepackage{lineno}
\usepackage[para,online,flushleft]{threeparttable}
\usepackage{float}
\RequirePackage{fontenc,inputenc,subcaption,graphicx,sectsty,titlesec,caption}
\RequirePackage{amsthm,amsmath,amssymb,animate,nccmath}
\RequirePackage{algorithm,paralist}
\RequirePackage[noend]{algpseudocode}
\definecolor{blue}{HTML}{40B4E6}
\definecolor{lightblue}{HTML}{BCBCF8}
\usepackage[colorlinks=true,linktoc=all,linkcolor=blue,citecolor=blue]{hyperref}
\usepackage[T1]{fontenc}
\usepackage{ae}
\usepackage[final]{microtype}
\usepackage{booktabs}
\usepackage{lineno}
\usepackage{textcomp}
\usepackage{setspace}

\newcolumntype{C}[1]{>{\centering\arraybackslash}m{#1}}

\def\b1{\mbox{$\mathbf{1}$}}

\newcommand{\betavec}{\boldsymbol{\beta}}

\definecolor{codegreen}{rgb}{0,0.6,0}
\definecolor{codegray}{rgb}{0.5,0.5,0.5}
\definecolor{codepurple}{rgb}{0.58,0,0.82}
\definecolor{backcolour}{rgb}{0.95,0.95,0.92}

\lstdefinestyle{mystyle}{
  backgroundcolor=\color{backcolour}, commentstyle=\color{codegreen},
  keywordstyle=\color{magenta},
  numberstyle=\tiny\color{codegray},
  stringstyle=\color{codepurple},
  basicstyle=\ttfamily\footnotesize,
  breakatwhitespace=false,
  breaklines=true,
  captionpos=b,
  keepspaces=true,
  numbers=left,
  numbersep=5pt,
  showspaces=false,
  showstringspaces=false,
  showtabs=false,
  tabsize=2
}

\title{{\Large{\bf Development of a Statistical Predictive Model  for Daily Water Table Depth and Important Variables Selection for Inference}}}
\author[1]{Alokesh Manna}
\author[2]{Sushant Mehan}
\author[3]{Devendra M. Amatya}
\affil[1]{Ph.D. Student in Department of Statistics, University of Connecticut, Department of Statistics,
Storrs, CT 06269, United States}
\affil[2]{Assistant Professor of Water Resources Engineering, Department of Agricultural and
Biosystems Engineering, South Dakota State University, Brookings, SD – 57006}
\affil[3]{Research Hydrologist. USDA Forest Service Santee Experimental Forest, Center for Forest Watershed Research, Cordesville, SC - 29434}

\begin{document}
\maketitle
\begin{center}



\begin{abstract} 

Accurately predicting water table dynamics is essential to sustain groundwater resources that support ecological functions, habitats, and anthropogenic activities. This study presents an assessment of a statistically driven model (BigVAR) for estimating water table depth using daily hydroclimatic variables USDA Forest Service Santee Experimental Forest Station on a low-gradient study site with high water table soils in Cordesville, SC (WS80, WS77, and WS78) and a site in NC (D1). The duration of data for the model development was from 2006 to 2019 and 1988 to 2008 for South Carolina and North Carolina, respectively. The variables used in the model were soil temperature, air temperature, precipitation, daily flow, wind speed, wind direction, relative humidity, solar radiation, net radiation, potential evapotranspiration, and groundwater well depth from multiple wells in South Carolina and North Carolina. Advanced statistical techniques, including regression analysis and data-driven modeling, were employed to construct predictive models that capture the intricate relationships between hydro climatological variables and water table depth using BigVAR (a recently developed model that takes into account modeling sparse vector autoregressions with exogenous variables or predictors). For the well on WS80, during the daily testing phase (2016 - 2020) and for the dormant season (11/01 – 03/31), the RMSE of water table depth was 14.94 cm and 10.09 cm, respectively. We also found no impact of flow (discharge) on estimating water table depth during the growing season (04/01 – 10/31) in the developed model where daily rainfall and lag of rainfall were included. At the daily time step, the coefficient of determination (R2) was .93 for the dry year (2019) with a total precipitation of 1380.91 mm and .96 for the wet year (2016) with a total rainfall of 1742.88 mm on the WS80 site in SC. The variable selection assessment showed that the most influential variables affecting the prediction of water table depths include the previous 4 day’s existing water table depth, and solar radiation, net radiation, rainfall, and wind direction, each using a lag structure from the previous two days, respectively for WS80. The developed predictive model can serve as a valuable tool for estimating water table depth, enabling forest managers, hydrologists, and regulatory agencies to make informed decisions regarding critical assessment of wetland hydrology in silvicultural management.

\vspace{0.1in}
\end{abstract}

\end{center}
\section{Introduction}

Water table depth plays a crucial role in silviculture management, influencing various ecological and hydrological processes essential for forest health and productivity. Understanding water table dynamics is vital for effective forest management practices, as it directly affects tree growth, species composition, and overall forest ecosystem resilience. Firstly, water table depth significantly impacts vegetation growth and health. Optimal water table levels are critical for sustaining plant biomass and root development. For instance, studies indicate that a water table depth of approximately 3 meters is ideal for preventing land desertification and promoting healthy vegetation growth, as it maintains a suitable water balance and mitigates soil salinization (\cite{cui2005role}). Additionally, research has shown that variations in water table depth can lead to substantial differences in root growth and biomass accumulation in species such as *Populus alba*, highlighting the importance of maintaining appropriate water levels for silvicultural success (\cite{imada2008water}).  

Moreover, the relationship between water table depth and forest management practices is evident in hydrology and nutrient cycling. Silvicultural activities, such as harvesting and thinning, can alter water table levels and affect hydrological processes. For example, studies have demonstrated significant changes in hydrology and nutrient concentrations on forest sites due to silvicultural operations, including a decrease in water table levels immediately following forest harvesting, which can impact water quantity and quality (\cite{ssegane2017calibration}; \cite{appelboom2006temperature}). Effective management practices that consider water table dynamics can help mitigate these impacts, ensuring that forest ecosystems remain resilient to disturbances such as drought and pest infestations (\cite{pinno2021opportunities}; \cite{del2014hydrology}). 

Furthermore, integrating water-oriented silvicultural practices is becoming increasingly important in adaptive forest management strategies, particularly in semiarid regions. These practices aim to balance the ecological needs of forests with human water use, thereby enhancing the sustainability of forest ecosystems (\cite{del2014hydrology}; \cite{manrique2015light}). By understanding the hydrological implications of silvicultural practices, forest managers can implement strategies that optimize water use while maintaining forest health and productivity. 

This is why estimating water table depth in forested ecosystems is essential for understanding hydrological dynamics and managing forest resources effectively. Various simulation methods have been developed to model water table depth, each with unique approaches and applications. This synthesis discusses several prominent simulation methods, highlighting their strengths and limitations. One of the widely used models for simulating daily water table depth is the DRAINMOD (\cite{tian2012drainmod}, \cite{grace2006hydrologic}, \cite{amatya2024hydrometeorological}). \cite{dai2010bi} describes another useful model MIKE SHE(originally developed by \cite{graham2005flexible}) which incorporates streamflow and water table depth to describe a hydrological process for a forested SC coastal plane. \cite{sun1998modeling} developed a model FLATWOOD for the forest hydrology of the wetland-upland ecosystem in Florida with daily precipitation and temperature. 

The integration of hydrological models with remote sensing techniques has also gained traction in estimating water table depth. For example, studies have demonstrated that satellite-derived data, such as the Normalized Difference Vegetation Index (NDVI), can be effectively used to assess water conservation and its relationship with water table levels in different ecosystems, including forests (\cite{zhang2021water}). \cite{zhang2021water} provides a spatially explicit approach to monitor water table changes over large areas, although the inherent variability of vegetation cover and other environmental factors may influence it. Furthermore, machine learning algorithms have emerged as a promising method for estimating water table depth. These algorithms can analyze vast datasets to identify patterns and relationships between various environmental variables and water table levels. For instance, explored the inherent water-use efficiency of different forest ecosystems, demonstrating how machine learning can enhance the understanding of water dynamics about climatic variables (\cite{liu2022inherent}). This approach allows for more accurate predictions of water table depth, particularly in the context of changing climate conditions. Lastly, the application of hydrological models that account for evapotranspiration processes is critical for estimating water table depth. Models that incorporate spatial evapotranspiration estimation methods have shown that inaccuracies in estimating this process can lead to significant uncertainties in hydrological simulations (\cite{yu2016hydrological}). One of the most prominent algorithms used for estimating water table depth is the Random Forest (RF) algorithm. This ensemble learning method has been shown to effectively model the depth to shallow groundwater at high spatial resolutions. For instance, demonstrated the applicability of RF in producing detailed maps that capture extreme conditions, such as wintertime water table depths, highlighting its robustness in handling non-linear relationships and interactions among predictors (\cite{koch2019modelling}). The RF algorithm's ability to manage large datasets and its resistance to overfitting make it particularly suitable for hydrological modeling in diverse forest environments. Support Vector Machines (SVM) have also been utilized to estimate water table depth. This method effectively classifies and regresses data, making it a valuable tool for predicting water table levels based on various input features. Additionally, machine learning techniques such as Artificial Neural Networks (ANNs) and Gradient Boosting (GB) have been applied in similar contexts. For example, explored various machine learning methods, including ANN and GB, for predicting water table depth in seasonal freezing-thawing areas. Their findings indicated that these methods could effectively capture the temporal fluctuations of water tables influenced by climatic and land-use changes (\cite{zhao2020machine}). The adaptability of these algorithms to different environmental conditions enhances their applicability in forest ecosystems. 

Moreover, the integration of multiple machine learning algorithms has been proposed to improve prediction accuracy. For instance, the combination of RF and other models can leverage the strengths of each method, resulting in more reliable estimates of water table depth.  

\begin{figure}[h!]%
    \centering
    \subfloat[\centering]{{\includegraphics[width=15cm,height=.45\textheight]{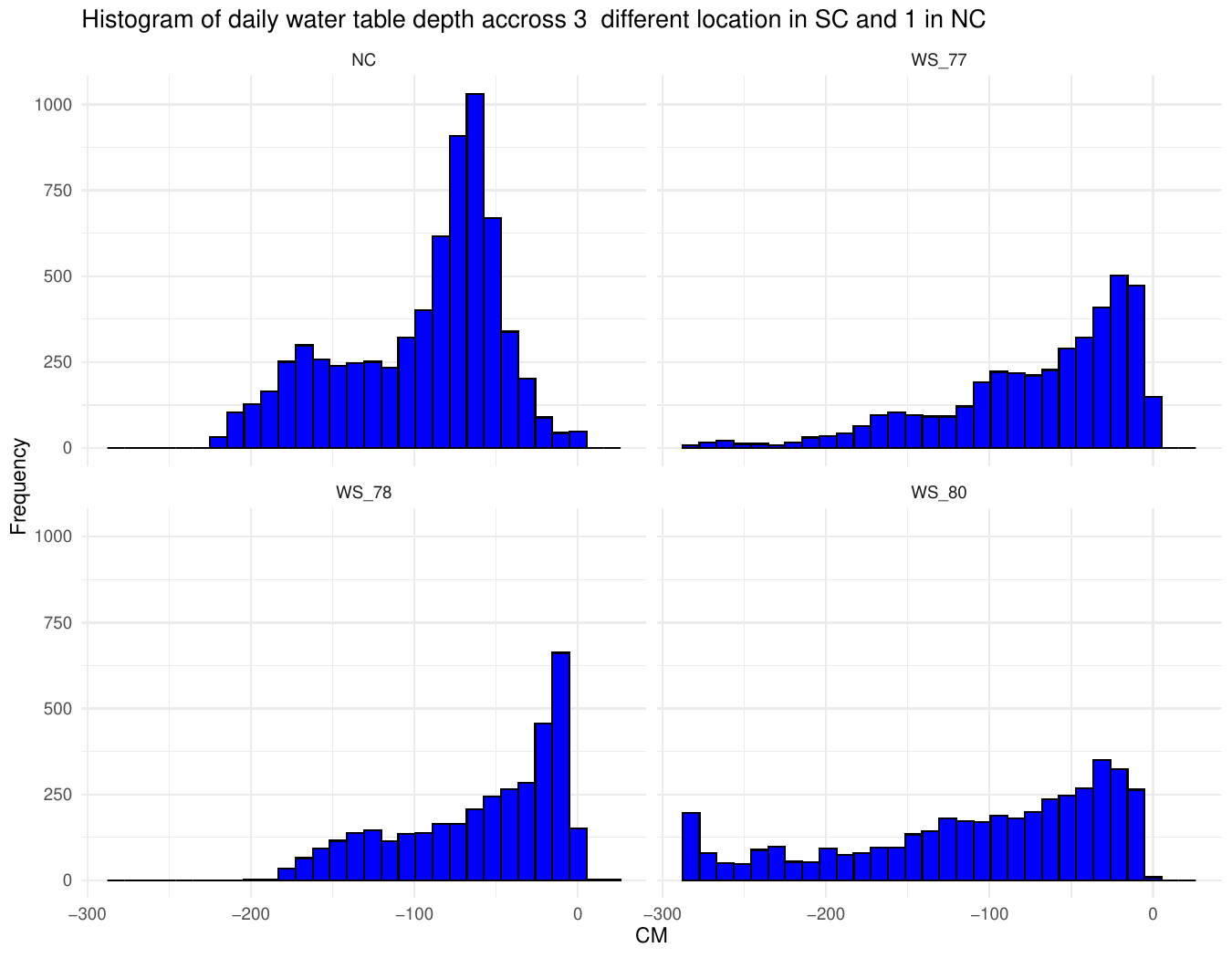} }}%
     \caption{Histogram of daily water table depth (cm) across three different well locations in watersheds WS77 and WS78 in SC and one in watershed D1 in NC}%
    \label{fig:hist_water_table_depth_daily}%
\end{figure}

\begin{figure}[h!]%
    \centering
    \subfloat[\centering ]{{\includegraphics[width=15cm,height=.45\textheight]{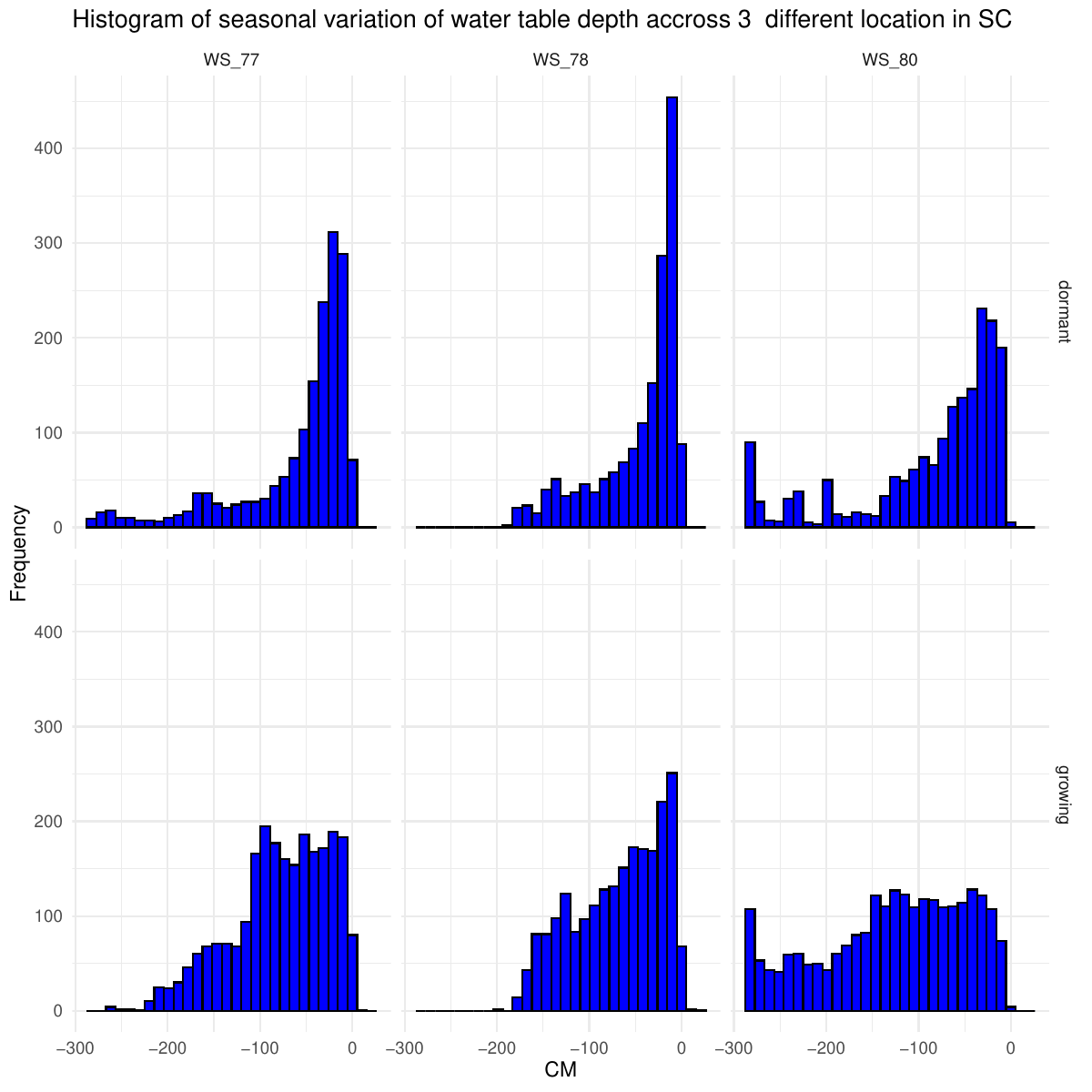} }}%
     \caption{ Histogram of daily water table depths (cm) for wells on watersheds WS77, WS78 and WS80 in SC for growing and dormant}%
    \label{fig:hist_water_table_depth_seasonal}%
\end{figure}

In this study, we will investigate the performance of a statistical model with optimization techniques to predict daily and seasonal (growing season (April 1 -to Oct 30) and dormant season (November 1 - March 31))  water table depth and elevation using available meteorological and hydrological data.  Weather and ecohydrology data, recorded at the USDA Forest Service Experimental Forest sites in South Carolina have been used to predict water table dynamics. Initially, we explored data from three hydro-meteorological stations across three watersheds: WS77, WS78, and WS80 (see figures \ref{fig:hist_water_table_depth_daily} and \ref{fig:hist_water_table_depth_seasonal}). In the figure \ref{fig:hist_water_table_depth_daily}, D1 (NC) has a relatively consistent and narrow distribution around -50 cm, indicating a stable water table depth around that range. WS77 shows a much narrower range and predominantly shallow water table depths, close to 0 cm. WS78 has a broader distribution with more variance in water table depths, extending to much deeper levels than WS77. WS80 shows multiple peaks, suggesting variability, with many occurrences at deeper water table depths, more so than at the other locations. In the figure \ref{fig:hist_water_table_depth_seasonal} we can observe that for dormant period the water table depths during the dormant period are relatively shallow, with most values concentrated near 0 cm for WS77 and WS78. The histogram shows a clear shift toward deeper depths during the dormant period, with frequent water table depths around -100 cm for WS80 in dormant season suggests that WS80 experiences deeper water tables compared to WS77 and WS78 during this season. for the growing period, the water table depths are much deeper, with frequent values between -100 cm and -200 cm for WS77, WS78, and WS80. The water table distribution is more spread out, indicating greater variability in water table depths during this period.

\begin{figure}%
    \centering
    \subfloat[\centering  TC met station]{{\includegraphics[width=3.5cm,height=0.15\textheight]{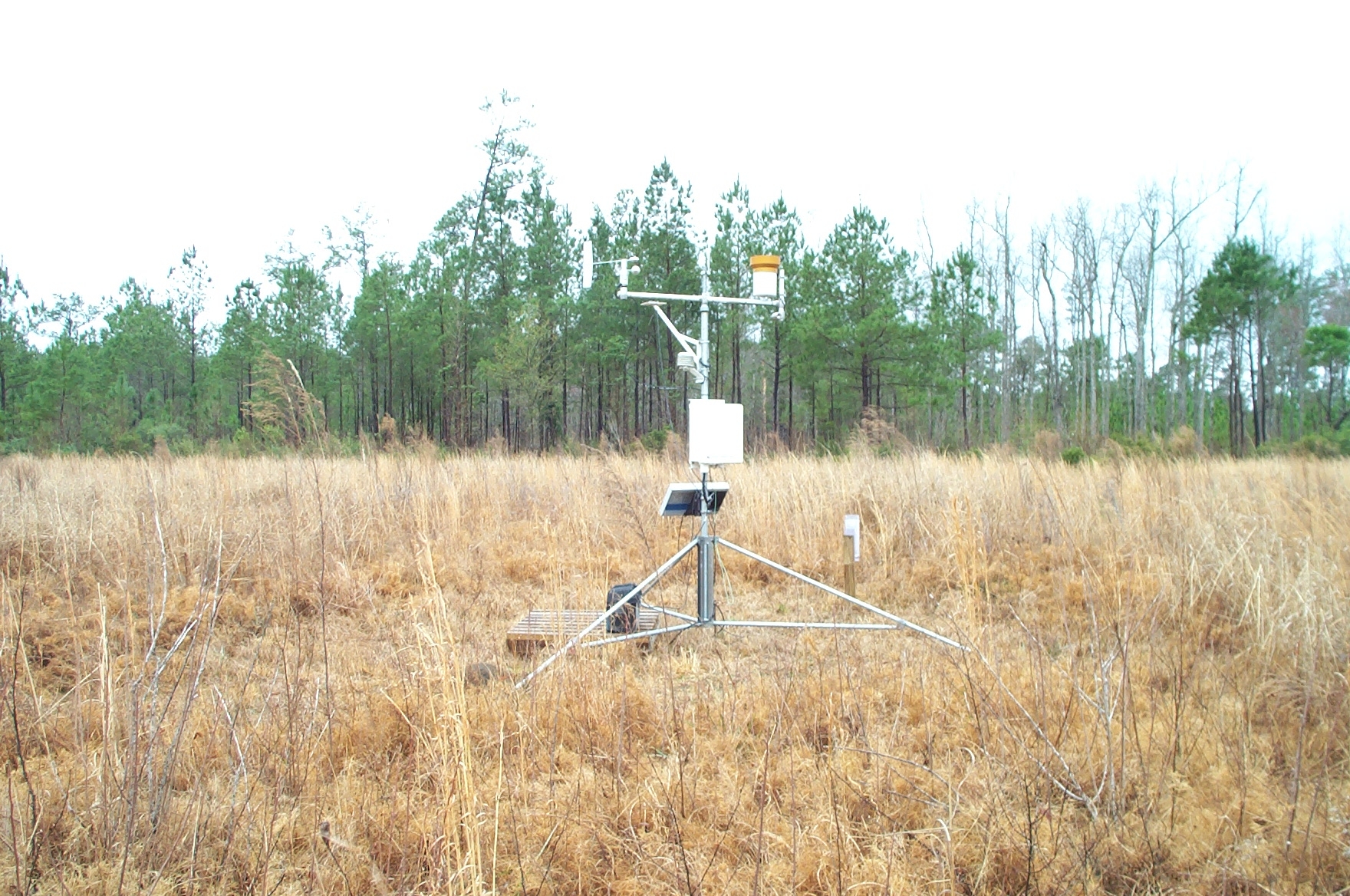} }}%
    \qquad
    \subfloat[\centering  WS78 rain gauge]{{\includegraphics[width=3.5cm,height=0.15\textheight]{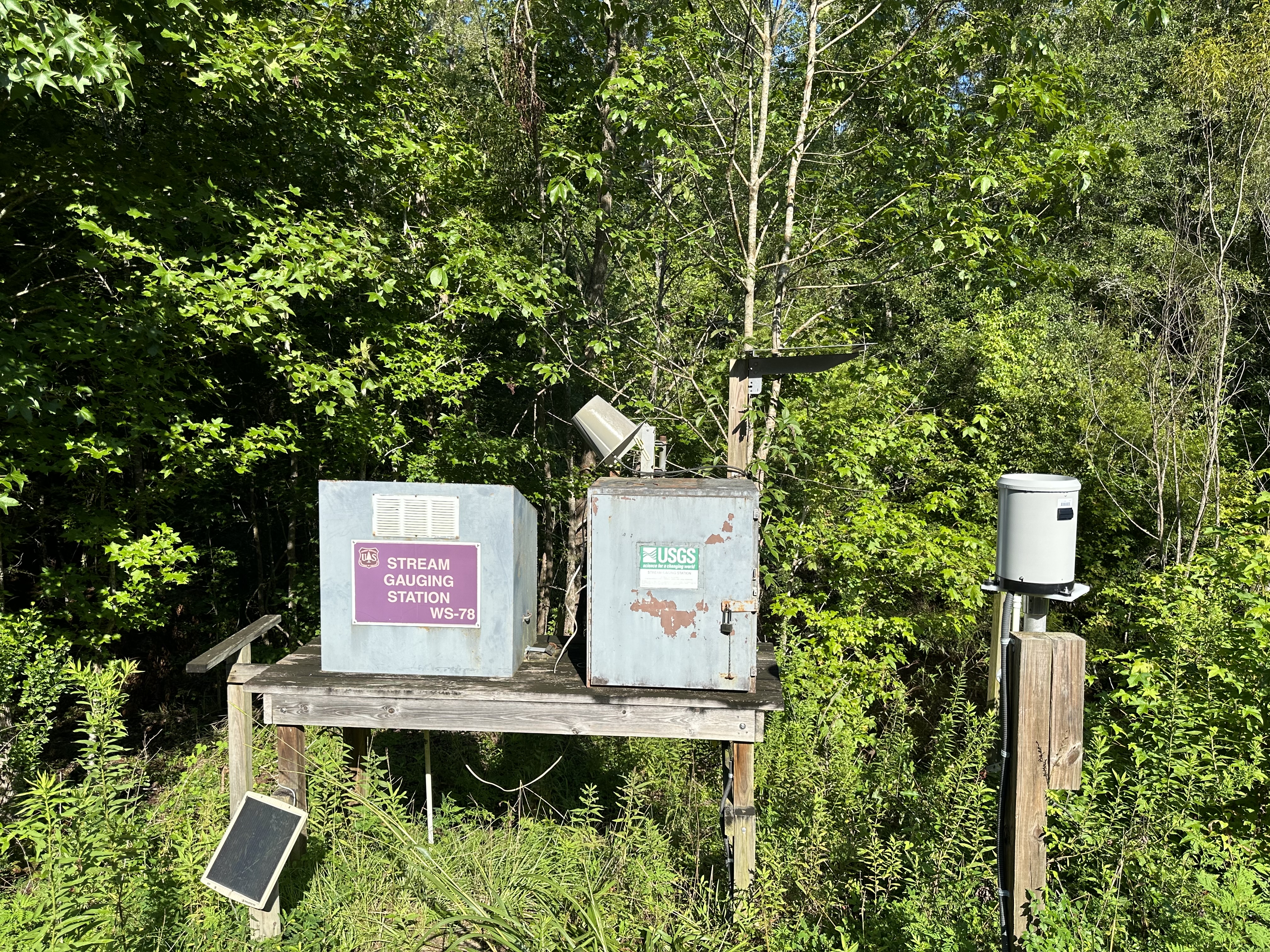} }}%
    \qquad
    \subfloat[\centering WS80 tower station]{{\includegraphics[width=3.5cm,height=0.15\textheight]{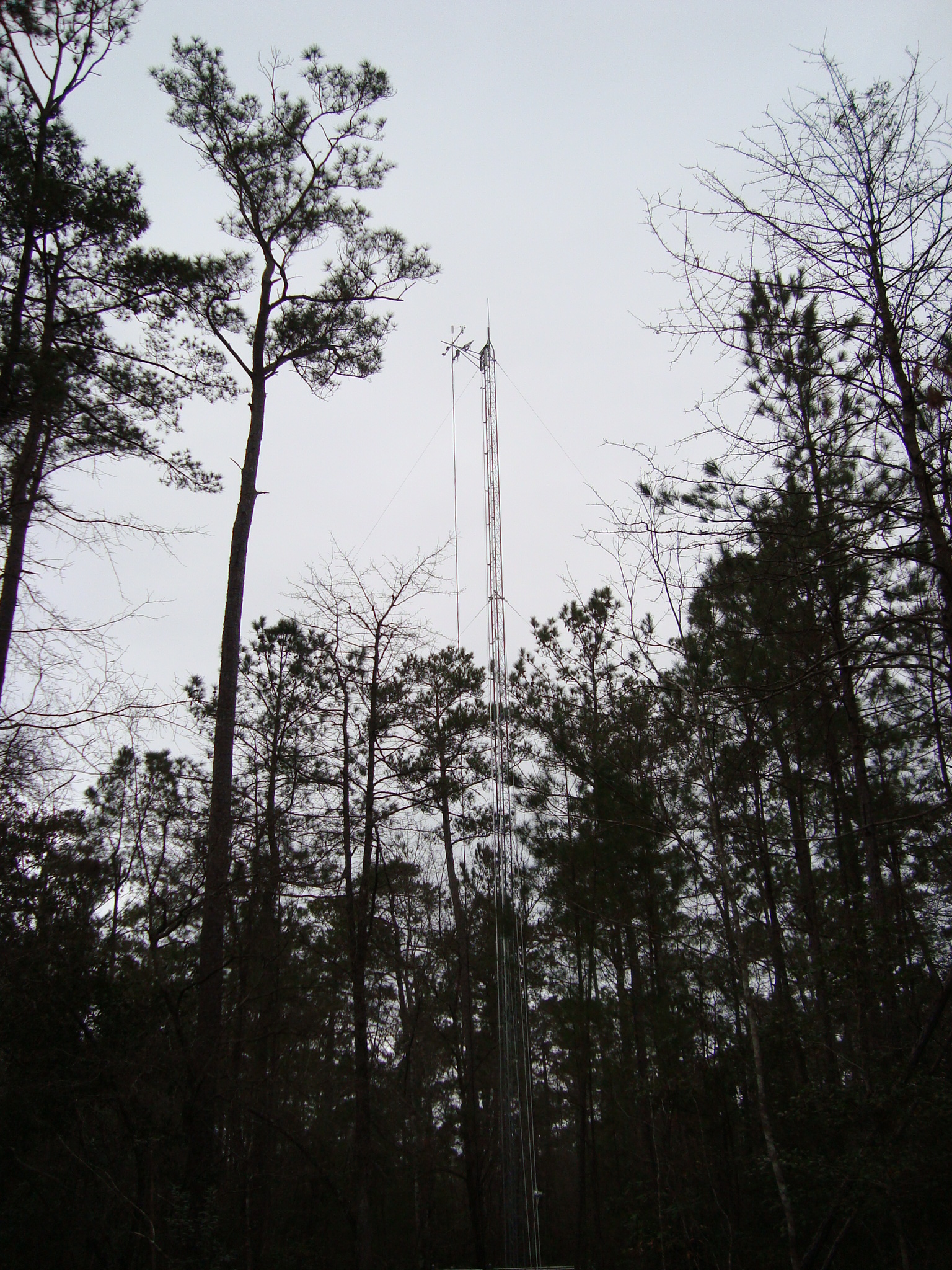} }}%
    \qquad
    \subfloat[\centering  SHQ station]{{\includegraphics[width=3.5cm,height=0.15\textheight]{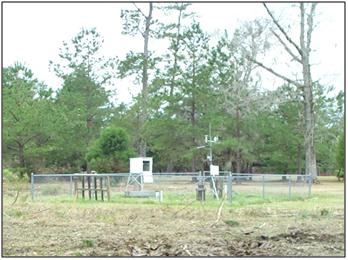} }}%
    \qquad
    \subfloat[\centering  Met 25 station]{{\includegraphics[width=3.5cm,height=0.15\textheight]{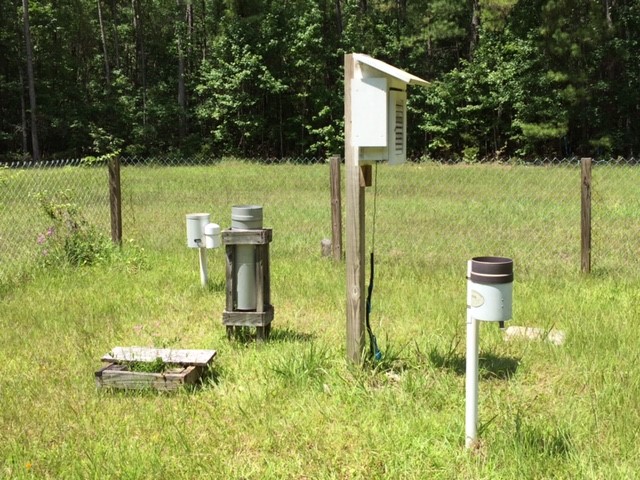} }}%
    \qquad
    \subfloat[\centering  Met 5 station]{{\includegraphics[width=3.5cm,height=0.15\textheight]{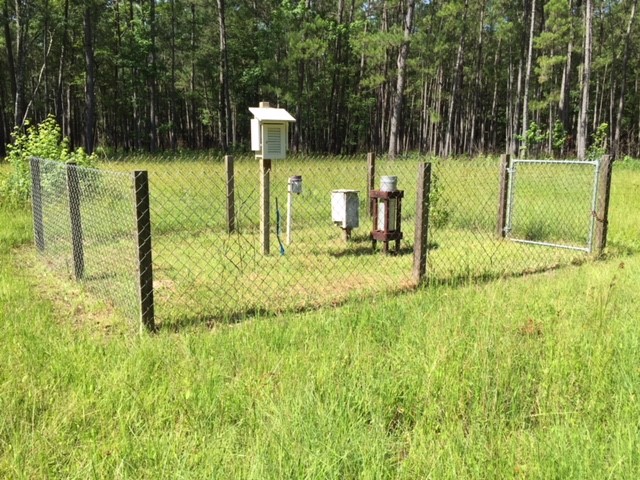} }}%
    \qquad
    \subfloat[\centering  Groundwater well]{{\includegraphics[width=3.5cm,height=0.15\textheight, angle =270]{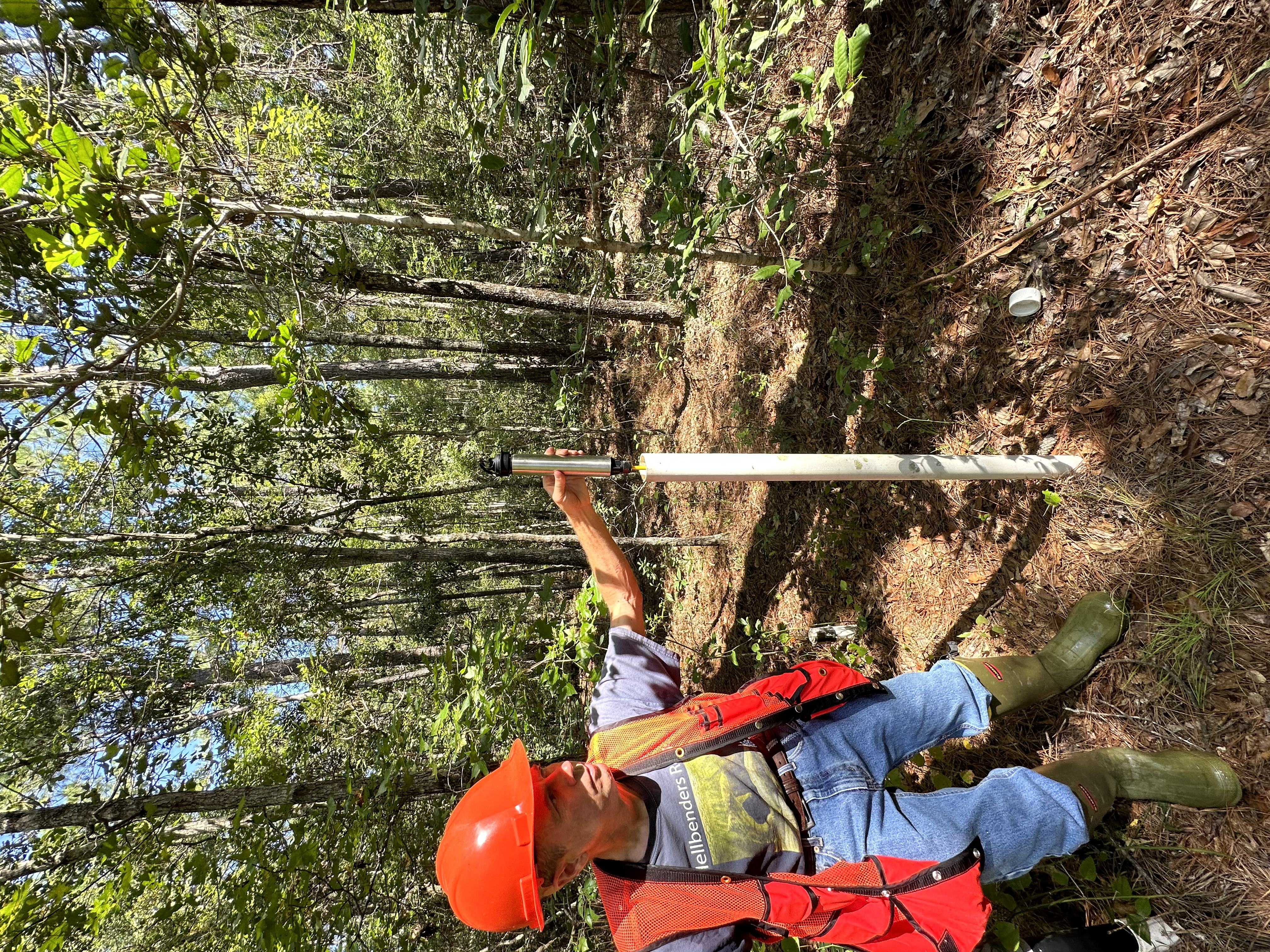} }}%
    \qquad
    \subfloat[\centering Flow measurement station]{{\includegraphics[width=3.5cm,height=0.15\textheight, angle =0]{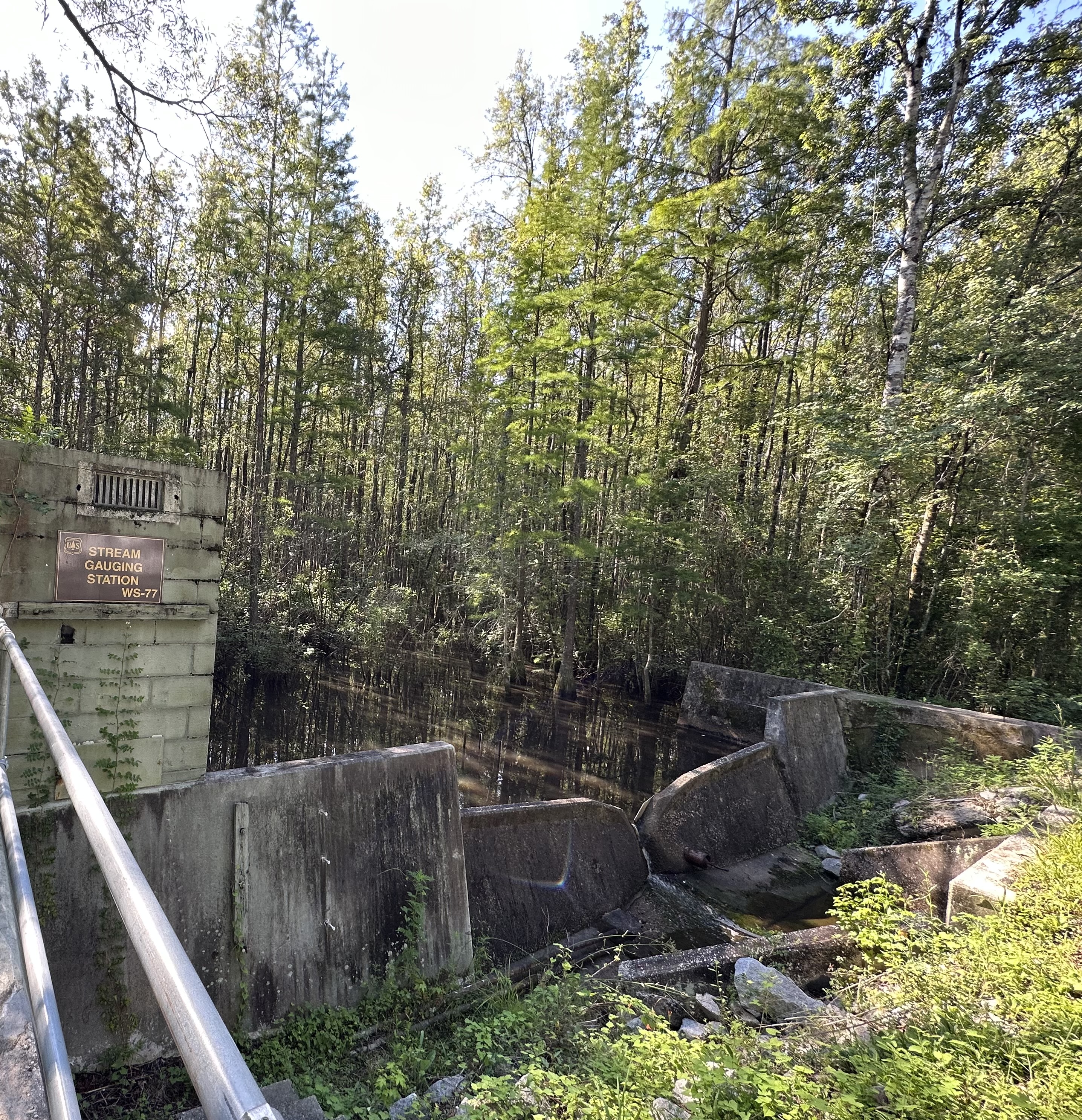} }}%
    \qquad
    \subfloat[\centering Recording logger in a gauge house]{{\includegraphics[width=3.5cm,height=0.15\textheight]{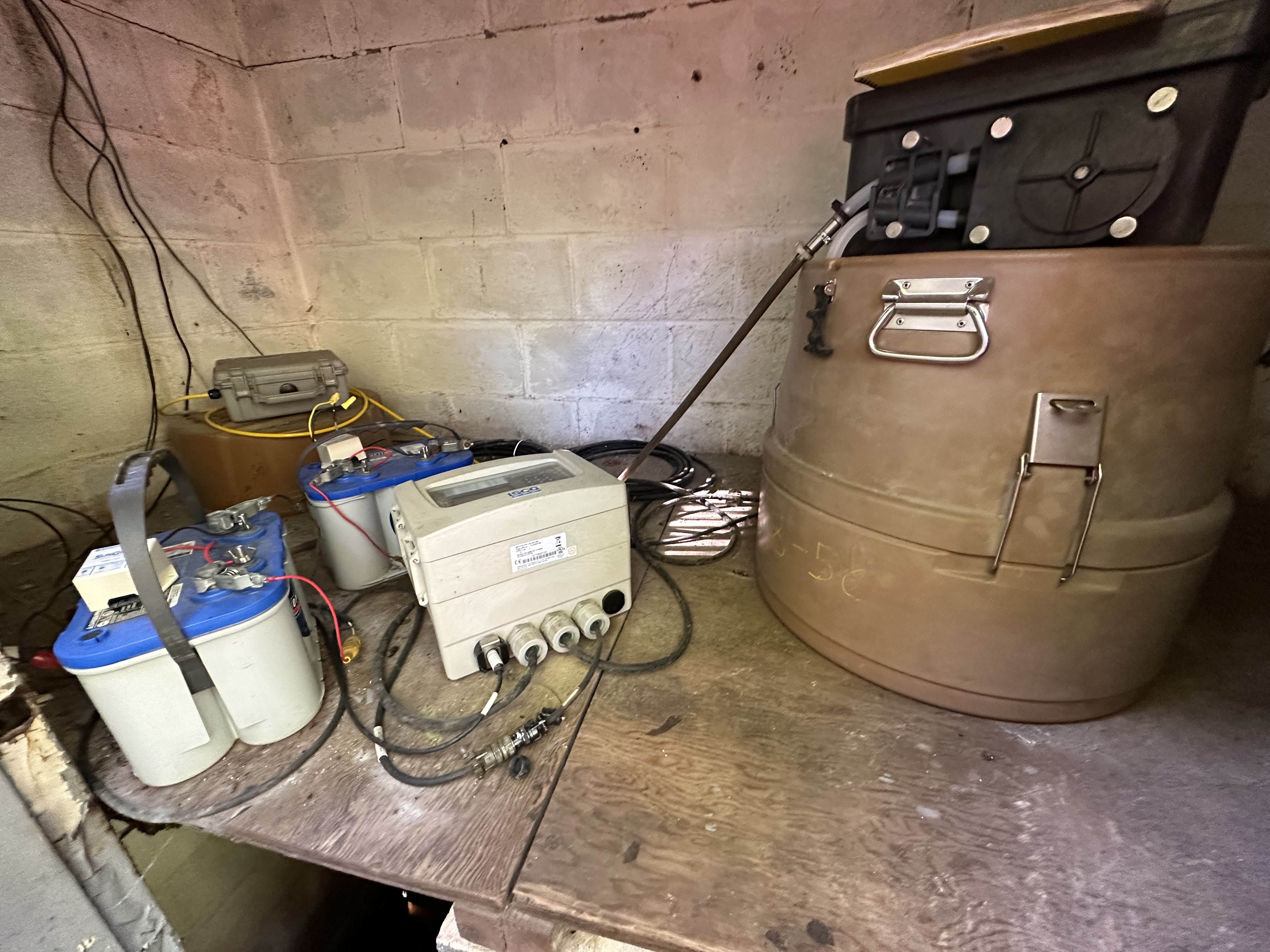} }}%
    \caption{Photos of various hydro-meteorologic measuring equipment on the study watersheds. Met 5 and Met 25 are satellite stations measuring air temperature, Soil temperature, and precipitation on WS77 and WS80, respectively. TC met is the complete weather station on the WS78 watershed. SHQ is another full-weather station in the headquarters office.}%
    \label{fig:watersheds}%
\end{figure}

\label{Literature survey} Water table depth in the soil is one important hydrologic variable that affects the growth of trees and biodiversity on the earth's surface. The water table, as used routinely by hydrologists in various disciplines, is seemingly a simple concept, that marks the top of the saturated zone in porous media (see \cite{baird2022water}).  The position of the water table relative to the ground surface— that is, water-table depth (WTD)—has also proved useful as an indicator of soil aeration and biochemical conditions of the soil. In addition, water table is also used to quantify and understand groundwater resources (\cite{baird2022water}). The authors also cited references indicating relationships between water table depth and many other variables, including soil greenhouse gas emissions, plant rooting depth, primary productivity, crop yield, and the species composition of wetland vegetation. Similarly, \cite{vepraskas2020method} used the water table depth to examine wetland hydrology on forested lands. These relationships exist because the water table is a proxy for the degree of waterlogging, soil redox status, and water availability. Therefore, an accurate prediction of water table dynamics is important. \cite{amatya2001hydrologic} discussed the water table depth prediction in pine plantations of poorly drained soils. It was concluded that the model is a reliable tool for assessing the hydrologic impacts of silvicultural and water management treatments, as well as climate changes on these pine stands. \cite{tian2012drainmod} has discussed the hydrology behavior in the presence of different chemical components in the soil and the effect on plant growth in drained soil. This application of DRAINMOD-FOREST demonstrated its capability for predicting hydrology and Carbon and Nitrogen dynamics in drained forests under limited silvicultural practices. \cite{amatya2019daily} discussed a methodology for water table depth prediction based on a differential equation with several available weather covariates such as rainfall, PET (potential evapotranspiration), etc. The proposed model predicts water table depth for poorly drained high water table soils, with the potential for assessing the effects of land management, wetland hydrology, and climate changes. Existing literature has addressed the water table depth estimation through the machine learning procedure. \cite{zhao2020machine} approached water table dynamics estimation with support vector regression. It was concluded that a water table depth of less than 3.64 billion m3 of water diversion might result in risks of environmental problems. \cite{herath2023deep} addressed the water level prediction model for the Colombo flood detention area using a feed-forward neural network and long short-term neural network. The LSTM has outperformed FFNN and confirmed that the temporal relationship is much more robust in predicting wetland water levels than the traditional relationship. \cite{nguyen2020combining} addressed the Red River water level forecasting problem with statistical and machine learning techniques. In this study, the time series data is considered in two parts. The linear part of the time series is handled by the autoregressive integrated moving average model(ARIMA) model, and different machine learning models, such as a random forest, handle the nonlinear part. \cite{zhu2020lake} have a review paper for lake water level forecasting using seven different machine learning models ANN(Artificial Neural Network), SVM (Support Vector Machine), ANFIS (Adaptive Neuro Fuzzy Inference System), hybrid models, evolutionary machine learning models, ELM (Extreme Learning Machine), and DL (Deep Learning). This paper presents a review of the applications of ML models for modeling water-level dynamics in lakes. \cite{ahmed2022water} discussed the problem of water level prediction using rainfall as a covariate with different methodologies such as linear regression, support vector regression, ensemble regression, XGboost, tree regression, and Gaussian process regression with a case study of Durian Tunggal river, Malaysia. Most machine learning algorithms try to fit the nonlinearity part but lose interpretability when the data diverges from linearity or many layers in neural network models. In water table dynamics, the value at a given time point is correlated with its previous values. Lag structures help capture these auto-correlations, stationarity, and causal effects allowing for better modeling and forecasting. If the data is very high dimensional with a small number of sample time points, \cite{manna2025some} can be used for modeling change points referring to moments in time when the statistical properties of hydrological data, such as river flow, precipitation, or groundwater levels, change abruptly. These changes might be due to natural events (like floods, and droughts), climate change, human activities (like land use changes, dam construction), or other environmental factors. The available hydro-climatic variables
are not also limited to affecting the water table depth at the same time point available but these variables observed several days before might have also influenced the current observed water table value. To address this problem, we have attempted to implement an auto-regressive model that incorporates lag structure on both the predictors and dependent variables and to perform the importance of the exogenous variables or predictors in the model. We have discussed the methodology below in section \ref{Methods}.
In our investigation, we sought to advance knowledge in modeling water table dynamics, which is critical in wetland hydrology assessment in silvicultural management. We attempted to accomplish that by selecting important variables for our inference among all available daily hydro-climatic variables for interpretability.  
We arranged our manuscript in the following way: in section \ref{Literature survey} we discussed a few available research articles for predicting water table dynamics. Section \ref{Research objective} described the research objective. After that, we discussed our data processing steps, implemented model, and a few exploratory data analyses in section \ref{Methods}. Next, we discussed our residual analysis in section \ref{Residual analysis} and important variables in section \ref{Variable selection}. Next, a discussion section \ref{conclusions_discussion} was provided on the justification of the consideration of this model from our end and collusive remarks. We presented the recommendation and future directions of our research in section \ref{Recommendations and future goals}. Lastly, we do include acknowledgement in section \ref{acknowledgement}.
\section{Research objective} \label{Research objective}
The key objectives of this research are the following: 
\begin{itemize}
    \item Identifying a statistical predictive model for evaluating water table depth measured at groundwater wells on three experimental watersheds WS77 , WS78, WS80 and one watershed D1 in NC across daily, monthly, growing, and dormant season temporal scales using daily hydrometeorological variables measured at the rain gauge, weather stations, and streamflow gauging stations within and/or in the vicinity of these groundwater wells as described below in Methods section. 
    \item Statistical inference for variable selection across different temporal scales and inference.
    \item Daily streamflow might not commonly available in different watersheds. So we investigate how the prediction changes when daily streamflow is not incorporated when we predict water table depth.
\end{itemize}

\section{Methods}\label{Methods}

\subsection{Site description}
The three study sites WS77, WS78, and WS80 located at the US Forest Service Santee Experimental Forest/Center for Forested Wetlands in \href{https://www.fs.usda.gov/rds/archive/Catalog/RDS-2019-0033}{wetland online available dataset}. The location diagram is given in figure \ref{fig:map}. For the user, one should go to this link, and in the ``Catalog/Data access:'' the zip folder of the data is available. The 155-ha headwater watershed WS77 was established in 1963 as a treatment in a paired system with WS80 (control) to study the hydrologic and water quality effects of prescribed burning on poorly drained coastal plain soils. A first-order stream drains WS77 into Fox Gulley Creek, eventually flowing into Turkey Creek and then Cooper River, which leads to the Atlantic Ocean. A few pictures from the site locations are provided in figure \ref{fig:watersheds}. 

Soils in WS77 belong mainly to the Wahee-Craven soil association, characterized by relatively poorly to moderately drained sandy loam to clayey soils with seasonally high water tables. The land is predominantly forested with loblolly pine, longleaf pine, and some bottomland hardwoods along the stream riparian bank. The watershed has low-gradient terrain with surface elevations ranging from 10.5 m to 5.6 m above mean sea level, with a topographic relief of up to 2\%\ slope. The region's climate is warm-temperate, with an average daily temperature of 16$^\circ$C and annual rainfall of 1375 mm, 40\%\ of which occurs during June-August. A gauging station with a compound concrete V-notch weir at the WS77 outlet measures stream outflows. For more details please check \cite{amoah2013quantifying}.

WS80 on the Santee Experimental Forest (SEF) was considered due to its diverse upland and wetland areas and long-term monitoring history. The 160-ha watershed WS80 serves as the control watershed for a paired system with WS77 within the larger second-order watershed WS79 (500 ha), which drains into Huger Creek, a tributary of the East Branch draining to the Cooper River. A gauging station with a compound concrete V-notch weir at the WS80 outlet measures stream outflows dating back to 1968.

The watershed's topography is planar with a slope of less than 4\%, and the elevation ranges from 4 to 10 m above mean sea level. Soils in WS80, developed from coastal plain sediments, are hydric and moderately well-drained in upland areas but poorly drained in the riparian zone. The main soil type is loamy, covering about 90\%\ of the watershed. Topsoil clay content is 30\%\ while subsoil clay content is 40-60\%\. The forest vegetation on WS80 has remained unregulated for over five decades, despite being heavily impacted by Hurricane Hugo in 1989. The site was left to regenerate naturally without biomass removal or salvage logging. A detailed description of this study site is available in \cite{dai2010bi}.

The third study site is the Turkey Creek watershed (WS78) in the Santee Experimental Forest. The WS78 gauging station has currently instrumented with a real-time stream gauge sensor and a rain gauge (rain gauge, accessed on 20 February 2024) since 2005 and is managed in cooperation with the US Geological Survey (USGS) and the College of Charleston, but it's monitoring has just been discontinued.

This watershed is a headwater of the East Branch of the Cooper River, which drains into Charleston Harbor. The Turkey Creek watershed, similar to other low-gradient forest watersheds in the southeastern Atlantic coastal plain, has experienced rapid urban development since the 1990s. The elevation ranges from 3.5 m at the outlet to 11.5 m above mean sea level.

The watershed's land use comprises 88\%\ pine forest (mainly regenerated loblolly and longleaf pine), 10\%\ wetlands and water, and 2\%\ agricultural lands, roads, and open areas. The forest was significantly impacted by Hurricane Hugo in 1989, leading to a mixture of remnant large trees, natural regeneration, and approximately 1000 ha of planted pine. Forest management practices include prescribed fire and thinning to reduce wildfire risks and support longleaf pine restoration and wildlife habitats, notably for the endangered red-cockaded woodpecker. For more details on WS78, please visit \cite{amatya2024hydrometeorological}.

The sub-tropical climate of above three study sites on the coastal plain features hot, humid summers and moderate winters. Winters are wet due to low evapotranspiration (ET) and long-duration rain events, while summers have high ET demands and short-duration, high-intensity storms, including tropical storms in July and October.

The fourth study site is located in Carteret County, North Carolina, and is managed by Weyerhaeuser Company. The location map is given in figure \ref{fig:map_nc}. Three artificially drained experimental watersheds (D1, D2, and D3), each about 25 ha in size, established in 1988 on a 14-year-old plantation. The site is flat with shallow water tables, and the soil is Deloss fine sandy loam, a hydric series. Each watershed is drained by four parallel lateral ditches, 1.4 to 1.8 m deep and spaced 100 m apart, which drain into a main roadside ditch via a collector ditch. The boundaries of the watersheds are defined by the mid-plane between the lateral ditches and rows of pine trees planted on 0.4 m high beds. Data on hydrology, soil, and vegetation parameters were collected from three experimental plots in each watershed. A detailed description of this watershed in NC can be found in \cite{amatya2011long}.

In our analysis, we focus on daily water table level prediction for all the above 4 study sites. We also conducted analyses of water table predictions on seasonal variations such as using monthly, growing, and dormant seasons for SC sites. While data from 2004 to 2021 were used for the groundwater wells at WS77 and WS80, 2006 to 2019 for the wells at WS78 in SC, data from 1988 to 2008 was used for the well at D1 at the NC site. Detailed descriptions of data monitoring techniques including sensors and loggers and their accuracy and limitations are provided by \cite{amatya2022long} for the SC sites and by \cite{amatya2011long} for the NC site. 

\subsection{Data Collection, Pre-processing, and EDA}
\begin{figure}[h!]

    \centering
    \includegraphics[width=0.7\textwidth]{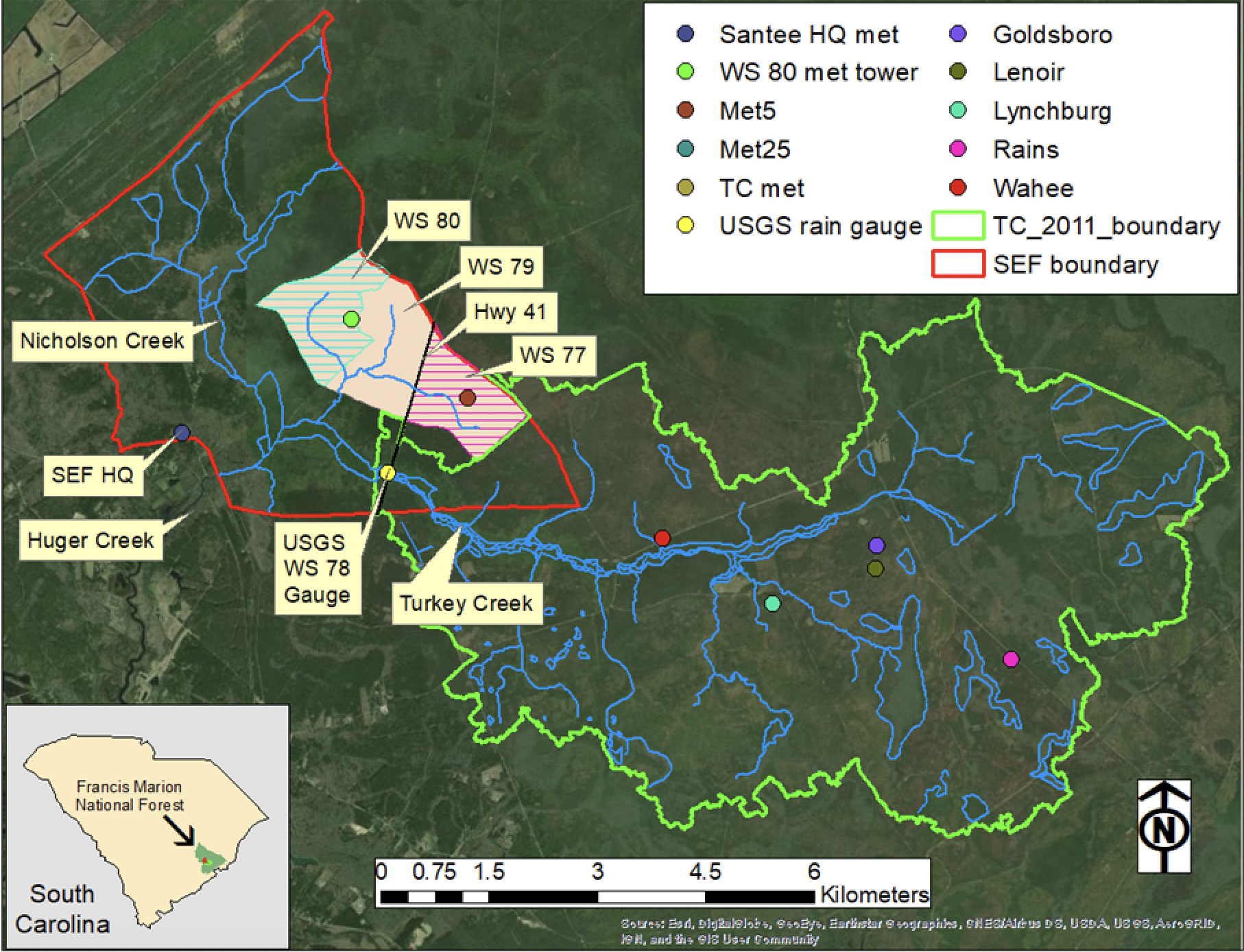}
    \caption{Location map of study watersheds WS77 and WS80 with Met5 and Met25, satellite stations, respectively, WS78 (Turkey Creek) in green boundary with TC met, a complete SEF HQ weather station, and the SEF HQ,  another full-weather station  at the Santee headquarters office.}
    \label{fig:map}
\end{figure}

\begin{figure}[h!]
\centering
    \includegraphics[width=0.8\textwidth]{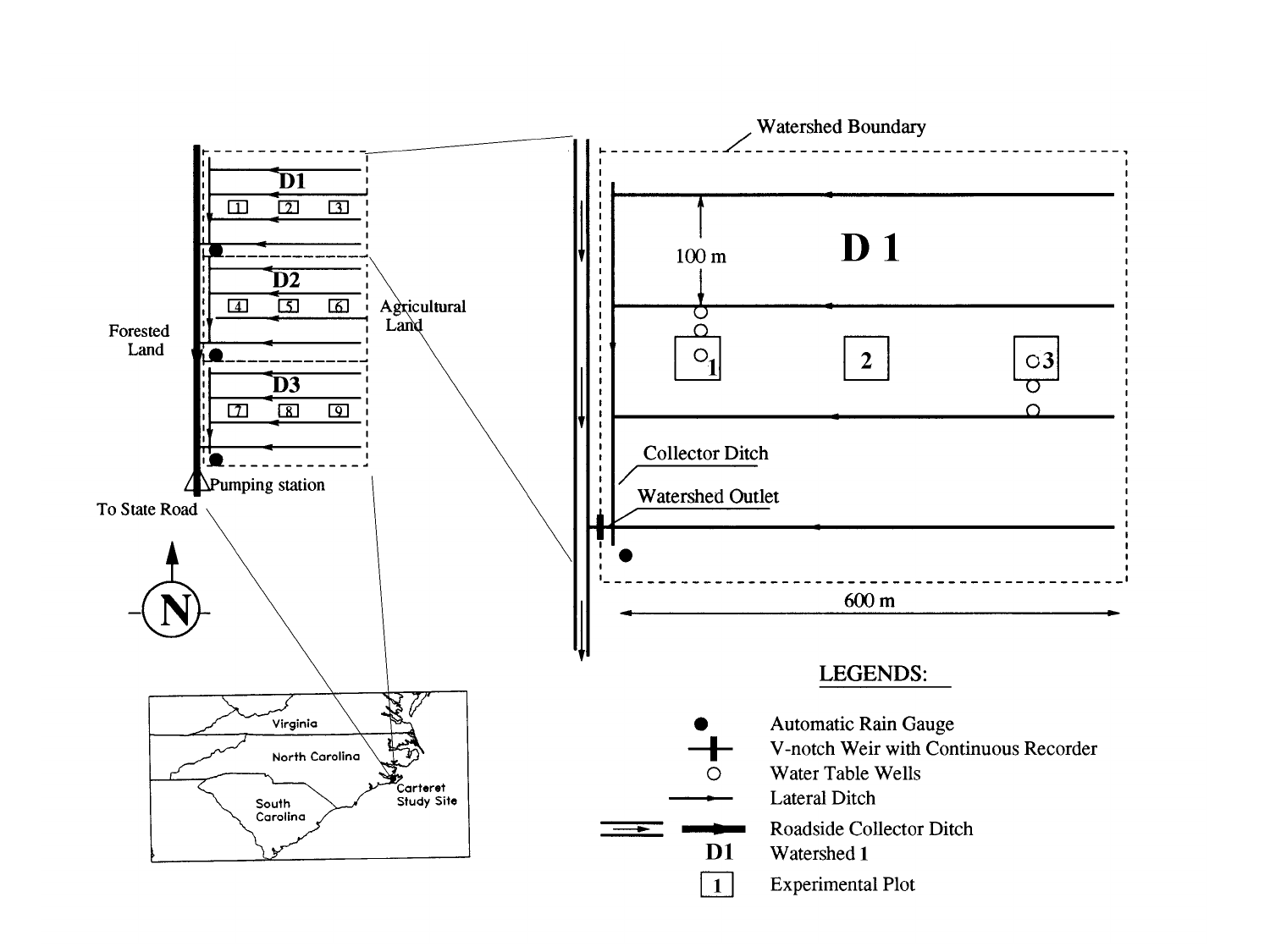}
    \caption{Location map of study watershed (D1) only, among two other adjacent watersheds D2 and D3, with hydrometeorological stations at the Carteret site in Coastal NC}
    \label{fig:map_nc}
\end{figure}

While systematically compiling all the data before processing, several pre-processing steps needed to be followed, as shown below.
\begin{itemize}
    \item There was missing data or data gaps for approximately 15 to 20 percent of all the data considered in the three watersheds. The data gaps can occur for different reasons. One of the reasons is power outrage in weather stations running with batteries. If the batteries are out of work, it might take a few days to get that reported by the authorized person for checking periodically. In our study, we have omitted the dates completely with all data, even if one of the variables is missing. We did not do any missing data imputation.
    \item WS77 did not have a full weather station with daily variables such as solar radiation, wind speed, RH, etc. So these data was borrowed from the nearest station at WS78. However, daily flow data, water table, air temperature, soil temperature, and precipitation were available for WS77.
    \item For WS80 study site, weather data is available from 2011 only. So the weather data was borrowed from another nearest weather station SHQ site (see figure \ref{fig:map}).
    \item Daily PET was estimated by the Priestley-Taylor method with the weather data
from WS80 and used for the other two stations assuming they do
not vary significantly in WS77 and WS78 sites. This is particularly because
daily net radiation data was not available to calculate daily PET at
each station.
    \item The net radiation for 2011-2019 was sometimes incomplete for the WS80 above
the tree canopy tower. The Forest Service performed the missing data imputation. For the 2006-2010 period at the TC station on WS78, the unavailable daily net radiation was estimated using a relationship between daily solar vs net radiation between the TC and WS80 stations.
    \item For WS78 and WS77 sites net radiation was calculated using the measured solar
radiation with the linear regression relationship between solar
and net radiation for the WS80.
     \item Units of the variables used: for streamflow - mm, water table depth - cm,
precipitation - mm, air and soil temperature - deg C, Relative
humidity - \%, solar and net radiation, Mj.sqm/day, Wind speed,
m/sec; wind direction - deg, and vapor pressure deficit - kPa.
\end{itemize}

\subsubsection{Exploratory data analysis}
\begin{figure}[h!]%
    \centering
    \subfloat[\centering WS 77]{{\includegraphics[width=7cm,height=0.35\textheight]{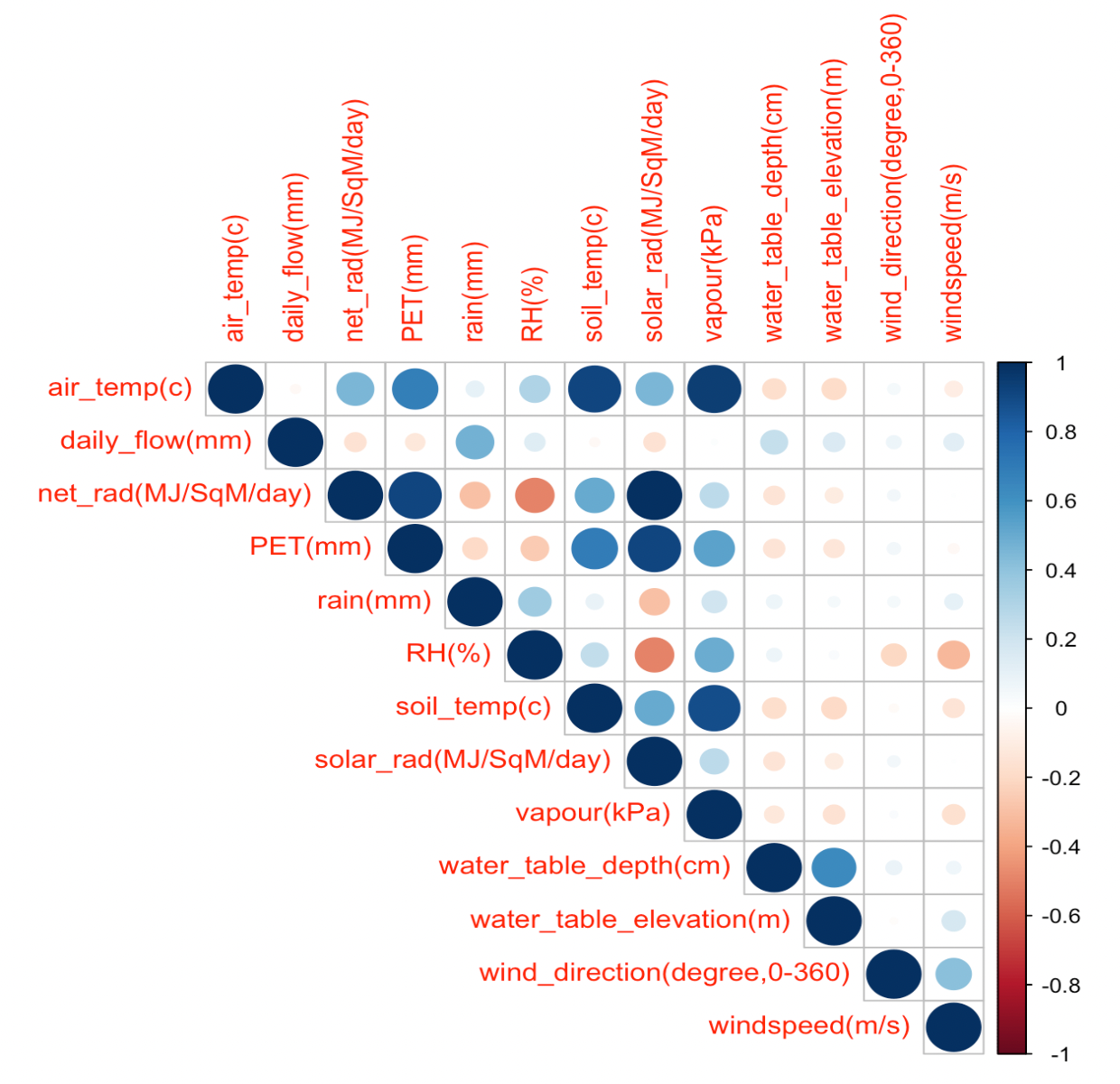 } }}%
    \qquad
    \subfloat[\centering  WS 78]{{\includegraphics[width=7cm,height=0.35\textheight]{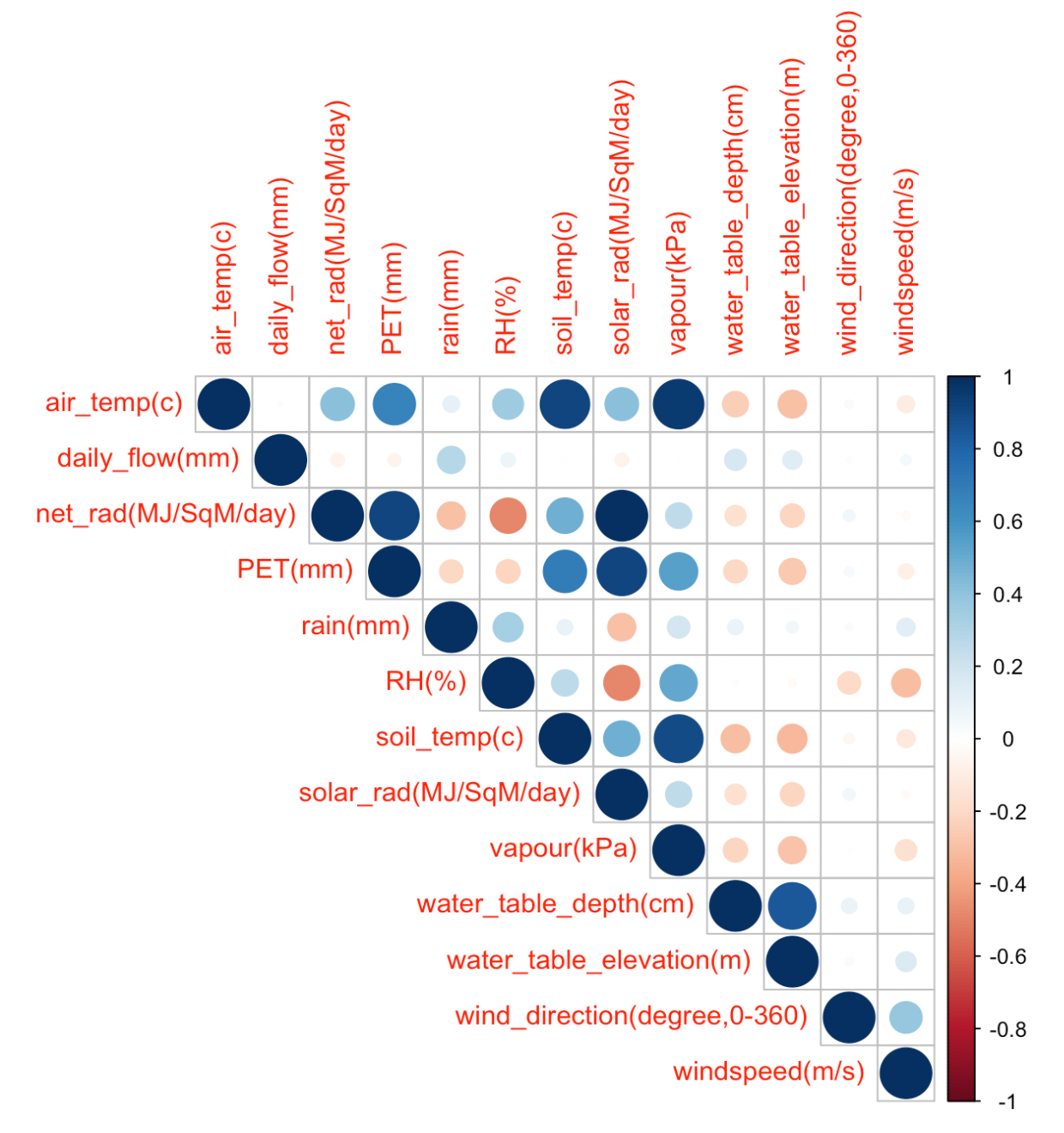} }}%
    \qquad
    \subfloat[\centering  WS 80]{{\includegraphics[width=7cm,height=0.35\textheight]{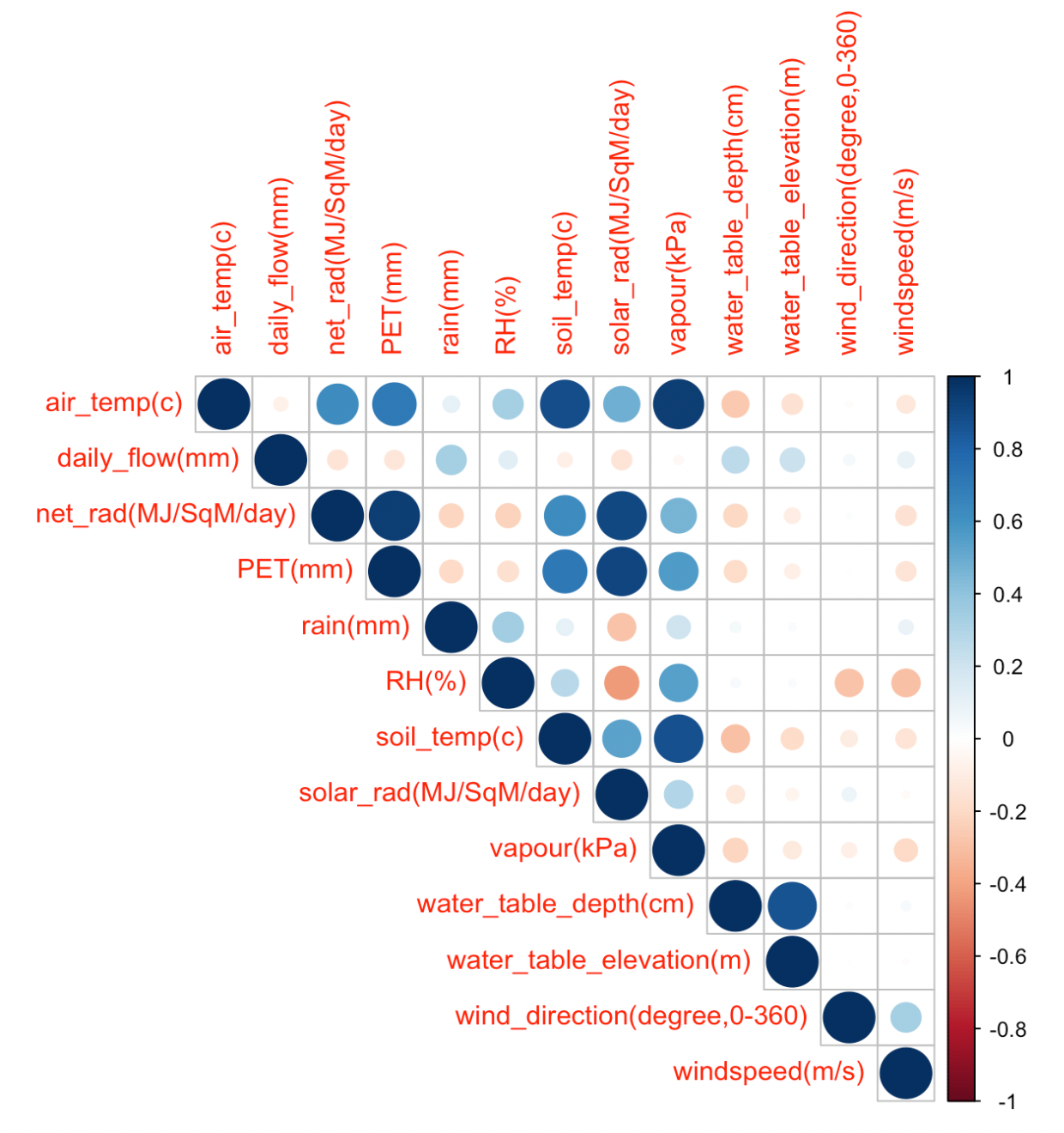} }}%
    \caption{Correlation plots for WS77, WS78 and WS80}%
    \label{correlation_plots}%
\end{figure}

Figure \ref{correlation_plots} explains the correlation plots as a matrix of all 13 variables measured on three watersheds.
The plot shows variables are intercorrelated with each other. For example, air temperature is correlated with potential evapotranspiration (PET), net radiation, solar radiation, soil temperature, vapor pressure, etc. Soil temperature and solar radiation are two closely correlated variables, so their intercorrelation patterns with other variables are similar. PET is another weather variable correlated with rain, relative humidity, soil and air temperature, vapor pressure, and water table depth. Water table depth is correlated with air temperature, daily flow, net radiation, PET, rain, relative humidity, soil temperature, and other covariates.

Water table elevation represents the height of the water table in the soil profile relative to a (fixed) datum, while water table depth (WTD) is the position of the water table in the soil profile relative to the ground surface at the recording well, are well correlated. Therefore, we focused on water table depth (WTD), which is more commonly used in hydrology analysis.

Based on the exploratory data analysis and correlation plots, we hypothesize that a combination of climatic variables available at nearby weather stations could explain daily water table depth and elevation dynamics (fluctuations in terms of magnitude, duration, and frequency) at a given site. The primary challenge is to build a model that can predict daily water table depth under this complex setup, where various hydro-climatic variables are non-uniformly correlated with each other.

\subsection{Final model selection and implementation}
A time series model that handles an auto-regressive process in time series in the presence of different covariates with their lag structure is required for our analysis. The model can be explained below.
\begin{align}
  \label{VAR1}
y_t=\mathbf{\nu}+\sum_{\ell=1}^p\Phi^{(\ell)}y_{t-\ell}+\sum_{j=1}^s \betavec^{(j)}\mathbf{x}_{t-j}+u_t \; \text{ for } \;t=1,\ldots,T.
\end{align}
In this model, predictors are represented as $\{\mathbf{x}_t\}_{t=1}^{T}$, which are daily hydro-climatic variables, such as soil temperature, air temperature, precipitation, daily flow, wind speed, wind direction, relative humidity, solar radiation, net radiation, and potential evapotranspiration. Groundwater well depth measurements $\{ \mathbf{y_t}\}_{t = 1}^T $obtained from multiple wells at the study site are called the water table depth. The water table depth is used as dependent variables $\{ \mathbf{y_t}\}_{t = 1}^T$ for our other model. $\betavec^{(j)}$ is the coefficient vector of $\mathbf{x}_{t-j}$, $j^{th}$ lag of the predictors and $\Phi^{(\ell)}$ is the $l^{th}$ coefficient for the lag structure of $y_{t}$. An additive random Gaussian noise $u_t\stackrel{\text{wn}}{\sim}(0,\sigma_{u}^{2})$ is considered in the model. \\
Let us define $\mathbf{\Phi}:=\left(\Phi^{(1)},\Phi^{(2)},\cdots,\Phi^{(p)}\right)$ as the vector of all coefficients of the lag structure of $y_{t}$ with order p and $\betavec:=\left(\betavec^{(1)},\betavec^{(2)},\cdots,\betavec^{(s)}\right)$ be the vector of coefficients of the predictors with their lag order s.
For the variable selection, this method uses penalized regression with the least square loss function of \ref{VAR1} as $\lambda\big(\alpha\|[\mathbf{\Phi},\betavec]\|_1+(1-\alpha)\|[\mathbf{\Phi},\betavec]\|_2^2\big)$, a method where all the coefficients are optimized over a smaller subset determined by $\lambda>0$.
The optimization function can be written as follows:
\[
\min_{\nu,\mathbf{\Phi},\betavec}\sum_{t=1}^{T_{1}}\|y_t-\mathbf{\nu}-\sum_{\ell=1}^p\Phi^{(\ell)}y_{t-\ell}-\sum_{j=1}^s \betavec^{(j)}\mathbf{x}_{t-j}\|^{2}+\lambda\big(\alpha\|[\mathbf{\Phi},\betavec]\|_1+(1-\alpha)\|[\mathbf{\Phi},\betavec]\|_2^2\big)
\]

$0\leq\alpha\leq 1$, is another parameter in the model which balances between the \(L_1\) and \(L_2\) norms of the coefficients. In general, for a vector \((x_1, x_2, \ldots, x_n)\), the \(L_2\) norm is calculated by taking the square root of the sum of the squares of all the components:
\[
\| \mathbf{x} \|_1 = |x_1| + |x_2| + \cdots + |x_n|,
\| \mathbf{x} \|_2 = \sqrt{x_1^2 + x_2^2 + \cdots + x_n^2}.
\]
\(L_1\) Norm is the distance you walk in a grid-like city (Manhattan distance). One adds up the absolute values of the differences in each direction. \(L_2\) Norm is the straight-line distance you would measure with a ruler (Euclidean distance). The Pythagorean theorem is used to calculate this distance. Both norms are useful in different contexts. The \(L_1\) norm is often used when you want to emphasize sparsity (making many values zero), while the \(L_2\) norm is used when the overall magnitude and smoothness are important factors to consider. Lasso tends to select only one variable from a group of highly correlated predictors, for example, in our case we can see in figure \ref{correlation_plots}, which can lead to instability in the selected model when multiple predictors are equally important. Ridge tends to shrink the coefficients of correlated predictors together, but it doesn't perform variable selection i.e., it doesn't zero out coefficients. Elastic Net strikes a balance between these two; it selects groups of correlated variables, unlike Lasso, and applies shrinkage like Ridge. This grouping effect is particularly useful when predictors are highly correlated, as it tends to select or exclude groups of related variables, leading to a more stable and interpretable model. Elastic Net allows for a weighted combination of both penalties. This flexibility can lead to better predictive performance, especially when neither Lasso nor Ridge alone is sufficient when there are many correlated features in high high-dimension setup.\\
The parameter $\lambda$ is selected by the grid search within an interval of 10 to 500. The complete time series data range is from 1 to $T$ which can be thought of as three segments. The first segment is 1 to $T_1$ which is commonly used as training data, the second segment is $T_{1}+1$ to $T_{2}$ which is used for parameter selection via cross-validation and the last segment is $T_{2}+1$ to $T_{3}$, a part of the time series of recent time, is used to evaluate the model performance. The period $T_{1}+1$ through $T_2$ is used to select $\lambda$. Define $\hat{y}_{\lambda, t+1}$ as the one-step ahead forecast based on $y_1, \dots, y_t$. $\lambda$ is chosen based on minimizing the one-step ahead mean square forecast error (MSFE) over the training period $T_{1}+1$ to $T_{2}$:
\[
\text{MSFE}(\lambda) = \frac{1}{T_2 - T_1 - 1} \sum_{t=T_1}^{T_2-1} \|\hat{y}_{\lambda, t+1} - y_{t+1}\|^2.
\] MSFE is calculated for a range of lambda values over a grid of values and the best lambda is chosen by cross-validation corresponding to minimum MSFE. $\alpha$ is a trade-off between the lasso (least absolute shrinkage and selection operator) and ridge penalties in the elastic net structure between 0 and 1. The default value for $\alpha$ is $\frac{1}{k+1}$ where $k=1$ in our case. It means we are taking equal weight of the lasso and ridge penalty in the penalization. The combination of the Lasso and ridge penalty is considered in this time series analysis method to improve the prediction accuracy for the ridge and interpretability of regression models for the lasso and ridge. The prediction of horizon 1 which is 1 day ahead of prediction in recent time is considered as testing data. $\sigma_{u}^{2}$ is unknown, so updated until convergence during the optimization procedure. \\
For the daily water table depth prediction, we used all the daily data for three of the watersheds in SC and D1 in NC. While doing the seasonal variants, for dormant season, we filtered the daily data only for the dormant season for each year and followed the similar procedure described above. For growing season, we filtered the daily level growing data for each of the year and followed similar modeling procedure. For the monthly prediction, we used the total amount of rainfall and average of remaining variables for each month of the years from the daily available watershed data. Although we do not recommend using the monthly average data for prediction in practice due to huge variability in different weather variables from their mean, we only show the performance in the tables. \\   
The testing period of this model $T_{1}+1$ through $T_2$ with the default $L_{2}$ loss and horizon as 1, an interval where $T_{1}=\lfloor\frac{\text{T}}{3}\rfloor$ and $T_{2}=\lfloor\frac{2\text{T}}{3}\rfloor$ and presented all the residual performance in the model which fits the least square. For the lag structure, we have used $p=4$ as the lag of water table depth and $s=2$ as the lag of different hydro-climatic variables that we use for inference as the maximum values. The values of $p$ and $s$ were selected based on the Bayesian Information Criterion (BIC), which combines the goodness of fit and the complexity into a single number. The Bayesian Information Criterion (BIC) is defined as:
\[
\text{BIC} = -2 \log(\hat{L}) + k \log(n)
\]
where $\hat{L}$ is the maximum likelihood of the model, $k$ is the number of parameters in the model, and $n$ is the number of data points. we compare our results based on the following metrics. Let \( y_i \) be the actual value, \( \hat{y}_i \) be the predicted value, \( \bar{y} \) be the mean of the actual values, \( n \) is the number of observations, and \( p \) is the number of predictors. Then the definitions of \( R^2 \), adjusted \( R^2 \), Mean Squared Error (MSE), and Root Mean Squared Error (RMSE) are defined below which we use as a comparison metric in different models and scenarios.
\begin{align*}
   & R^2 = 1 - \frac{\sum_{i=1}^{n} (y_i - \hat{y}_i)^2}{\sum_{i=1}^{n} (y_i - \bar{y})^2}\\
   &\text{adjusted } R^2 = 1 - \frac{(1 - R^2)(n - 1)}{n - p - 1}\\
   &\text{MSE} = \frac{1}{n} \sum_{i=1}^{n} (y_i - \hat{y}_i)^2\\
  &\text{RMSE} = \sqrt{\frac{1}{n} \sum_{i=1}^{n} (y_i - \hat{y}_i)^2}
\end{align*}
The Nash–Sutcliffe efficiency is equivalent to the coefficient of determination ($R^2$), thus ranging between $-\infty$ and 1. Mean Error (ME) measures the bias of the model; how much the average predicted value deviates from the observed. Can be positive or negative, indicating over- or under-prediction. Mean Absolute Error (MAE) is the average absolute difference between predicted and observed values. It is easier to interpret since it's in the same units as the data. Mean Squared Error (MSE) is the average of squared differences between predicted and observed values, emphasizing larger errors due to the squaring. Root Mean Square Error (RMSE) is the square root of MSE; in the same units as the data, emphasizing larger errors more than MAE. Unbiased Root Mean Square Error (ubRMSE) is similar with RMSE, but without bias. It isolates random error by removing systematic error (bias). Normalized RMSE (NRMSE) is normalized by the range or mean of the observed data, allowing comparison across different datasets. Percent Bias (PBIAS) shows the average tendency of predictions to over or under predict. A negative value indicates model overestimation. Ratio of RMSE to Standard Deviation (RSR) is a ratio of RMSE to the standard deviation of observed data. Low RSR indicates better model performance. Ratio of Standard Deviations (rSD) compares variability between predicted and observed values. A value near 1 indicates similar variability. The lower limit of NSE is $-\infty$, so to eliminate this problem $\text{NNSE}:=\frac{1}{2-\text{NSE}}$. Modified NSE (mNSE) modifies NSE to address issues with extreme values. Relative NSE (rNSE) considers relative differences between predictions and observations. Weighted NSE (wNSE) is a NSE variant that gives different weights to data points, often used for hydrological predictions. Weighted Seasonal NSE (wsNSE) is another NSE modified to account for seasonal effects, useful for predicting seasonal dynamics. Index of Agreement (d) measures agreement between predictions and observations. It ranges from 0 to 1, with 1 indicating perfect agreement. Refined Index of Agreement (dr) addresses the limitations of the original index of agreement. Modified Index of Agreement (md) is a variation of the index of agreement that better addresses certain model behaviors. Relative Index of Agreement (rd) is a relative measure of agreement between predictions and observations. Persistence Index (cp) is a measure that evaluates how persistent the predicted signal is compared to the observed. Pearson Correlation Coefficient (r) measures the linear relationship between predicted and observed values. ``r'' range from -1 (perfect negative correlation) to 1 (perfect positive correlation). Coefficient of Determination ($R^2$) represents the proportion of variance explained by the model. A value closer to 1 indicates a good fit. Volumetric Efficiency (VE) evaluates how well the model reproduces total volume of observed data, often used in hydrological models. Kling-Gupta Efficiency (KGE) combines correlation, bias, and variability to give a more holistic measure of model performance than NSE. KGE for Low Values (KGElf) is a version of KGE that gives more weight to low values in predictions, useful in low-flow regimes. Non-parametric KGE (KGEnp) is A non-parametric version of KGE, often more robust against outliers. We have applied the above metric to evaluate the performance of our model.

\subsection{Software availability}
``BigVAR'' (\cite{nicholson2017bigvar}) is an existing R Package designed to handle high-dimensional multivariate time series modeling using vector autoregressions (VAR). It provides efficient methods for estimating VAR models when the number of variables is large relative to the number of periods, which is common in many modern applications. The package focuses on overcoming the challenges of overfitting in such high-dimensional settings by incorporating various regularization techniques. The R package is available for this model (\cite{nicholson2019bigvar}). For further references see \href{http://www.wbnicholson.com/BigVAR.html}{BigVAR implementation in R}. Our codes for analysis and application are available in \href{https://github.com/alokesh17/Water_Table_Depth-.git}{GitHub}.

\section{Results}
\label{Residual analysis}

\subsection{Time series plots}
\begin{figure}%
    \centering
    \subfloat[\centering ]{{\includegraphics[width=16cm,height=.4\textheight]{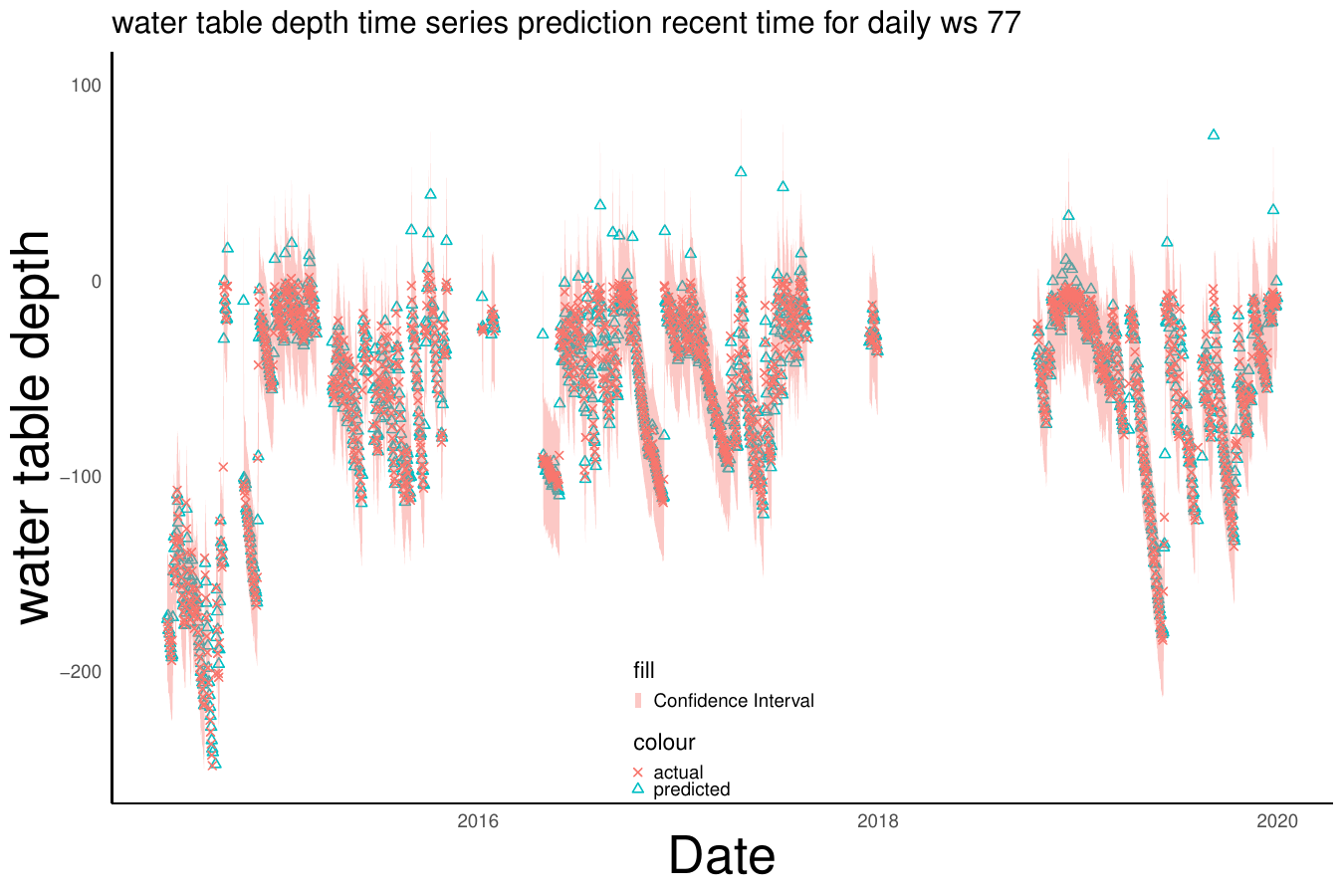} }}%
    \qquad
    \subfloat[\centering  ]{{\includegraphics[width=16cm,height=0.4\textheight]{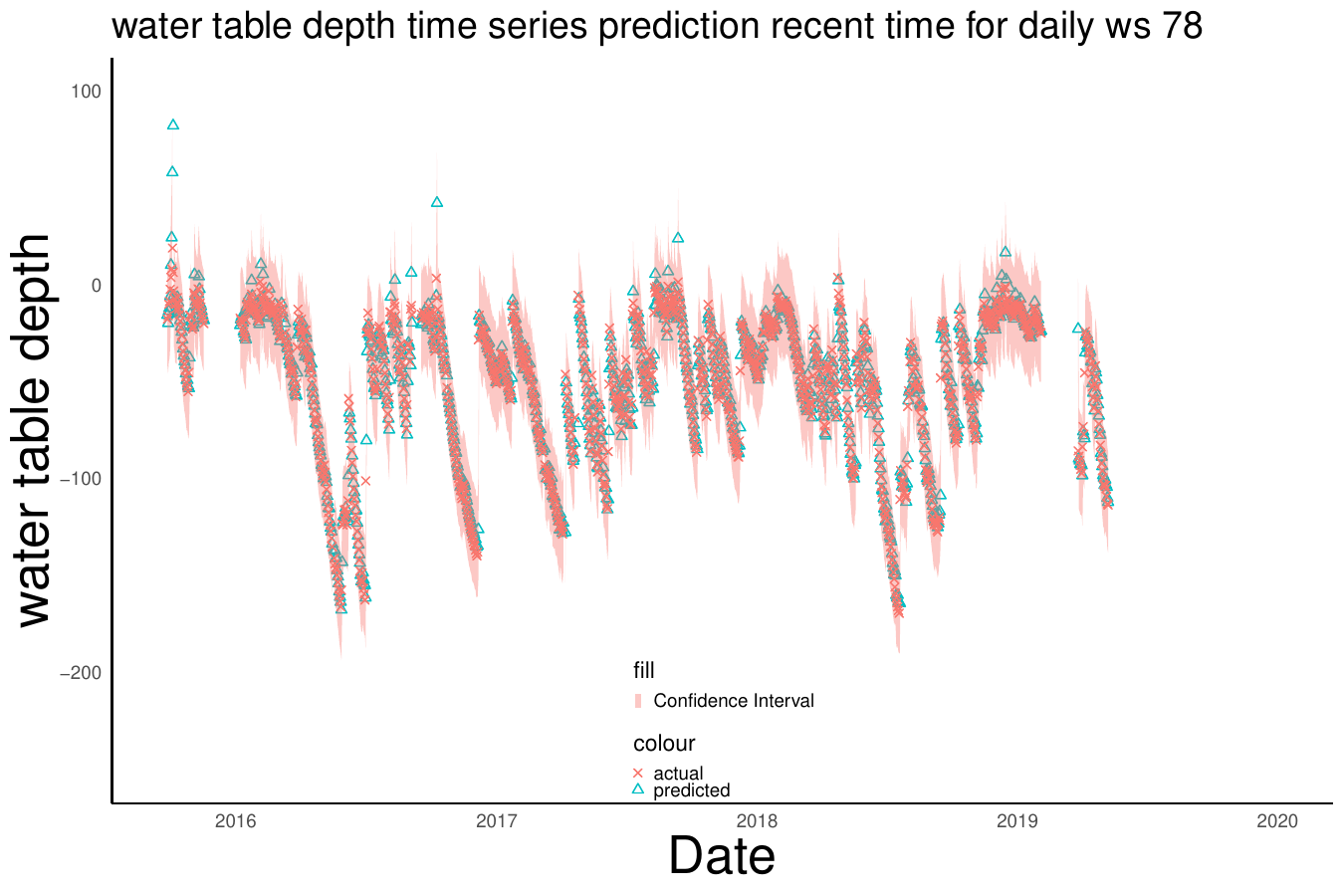} }}%
     \caption{ Time series of Predicted versus measured daily water table depths including their 95\% conﬁdence intervals (hatched area) for the wells (a) at WS 77 and (b) at WS 78 for the 2016-2019 testing period. }%
    \label{fig:ts_plots_77_78}%
\end{figure}

\begin{figure}%
    \centering
    \subfloat[\centering]{{\includegraphics[width=16cm,height=0.4\textheight]{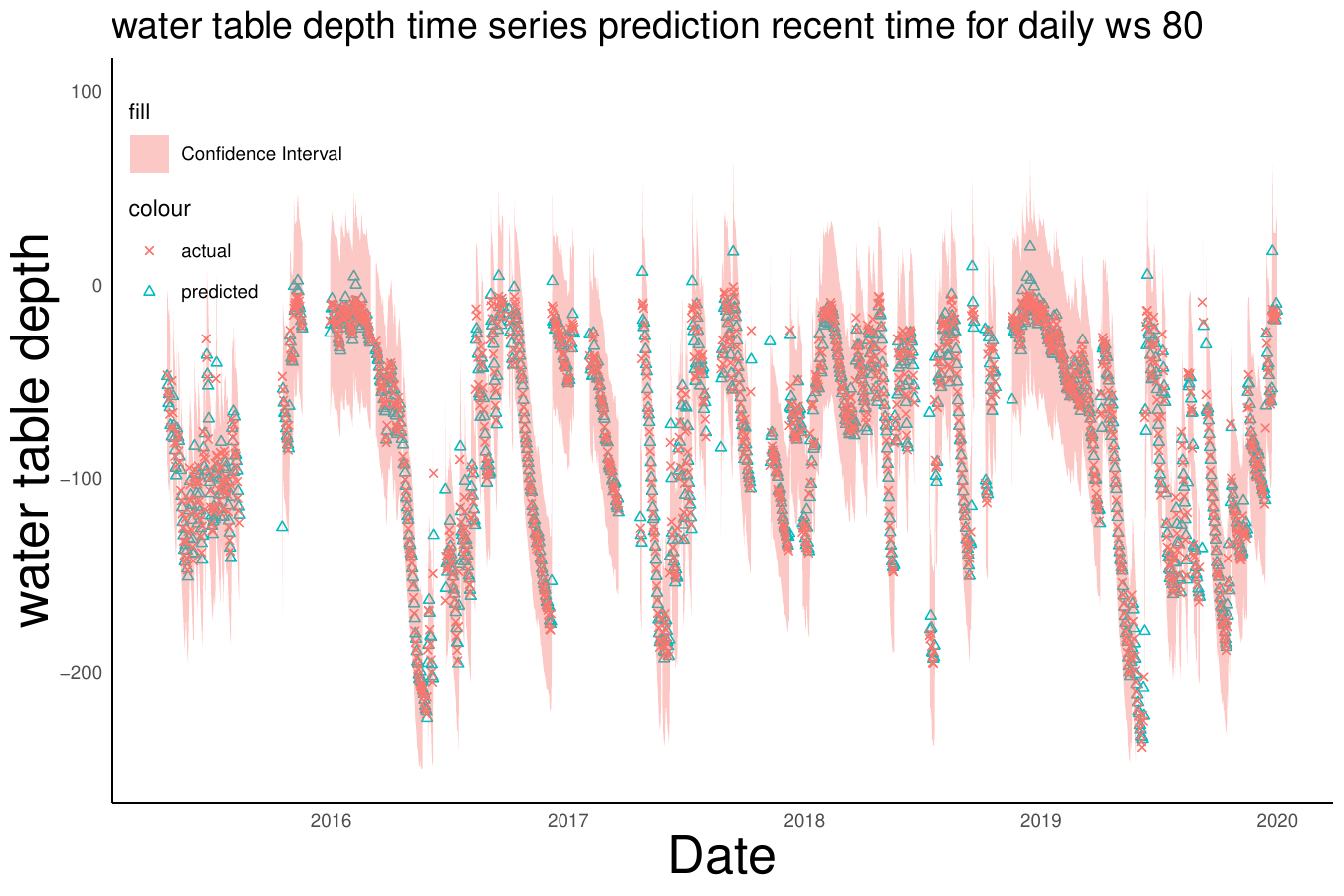} }}%
    \qquad
    \subfloat[\centering]{{\includegraphics[width=16cm,height=0.4\textheight]{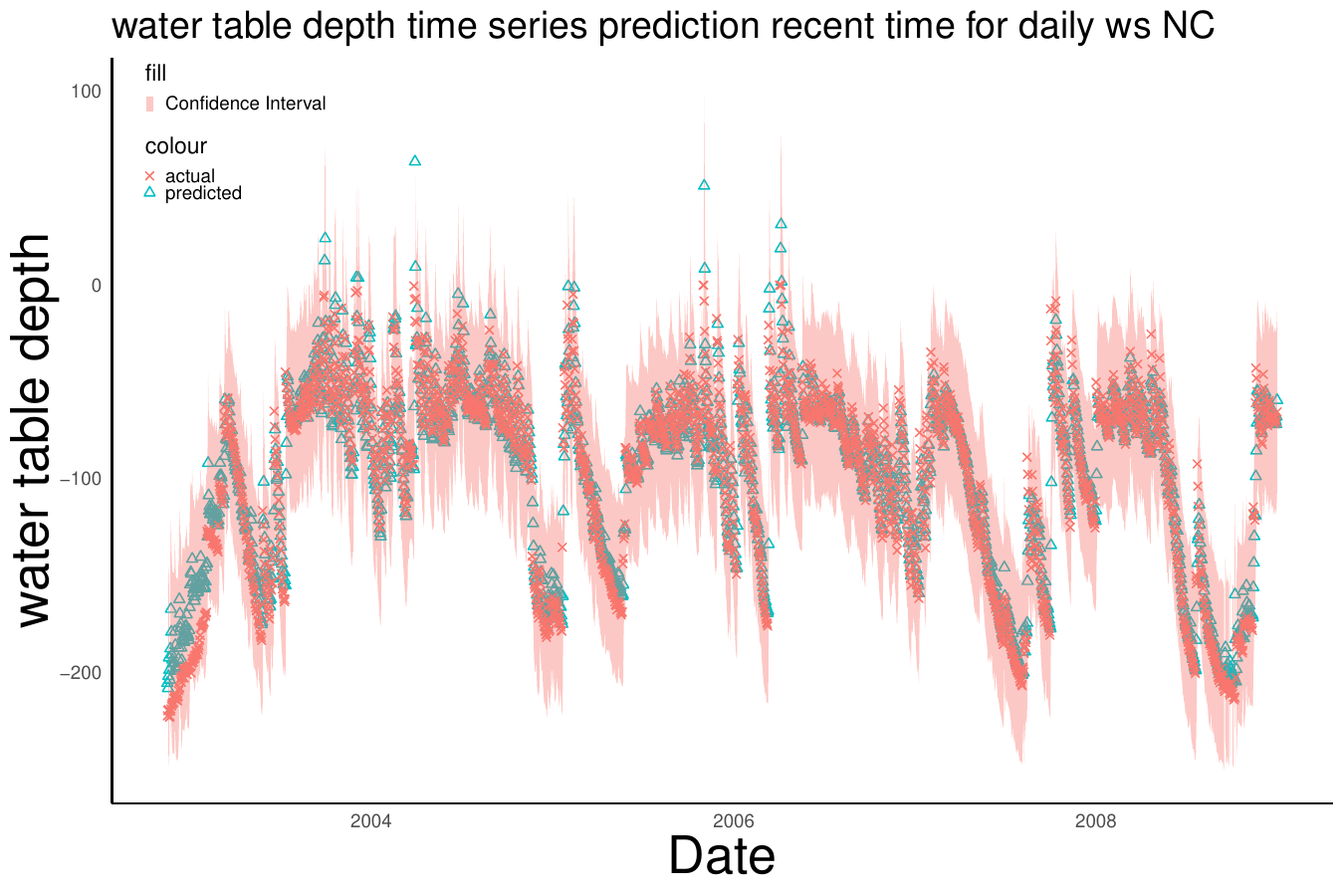} }}%
    \caption{
    Time series of Predicted versus measured daily water table depths including their 95\% conﬁdence intervals (hatched area) for the wells (a) at WS 80 for the 2016-2019 and (b) D1 in NC for the 2004-2008 testing period.}%
    \label{fig:ts_plots_80_NC}%
\end{figure}

\begin{figure}%
    \centering
    \subfloat[\centering ]{{\includegraphics[width=16cm,height=.4\textheight]{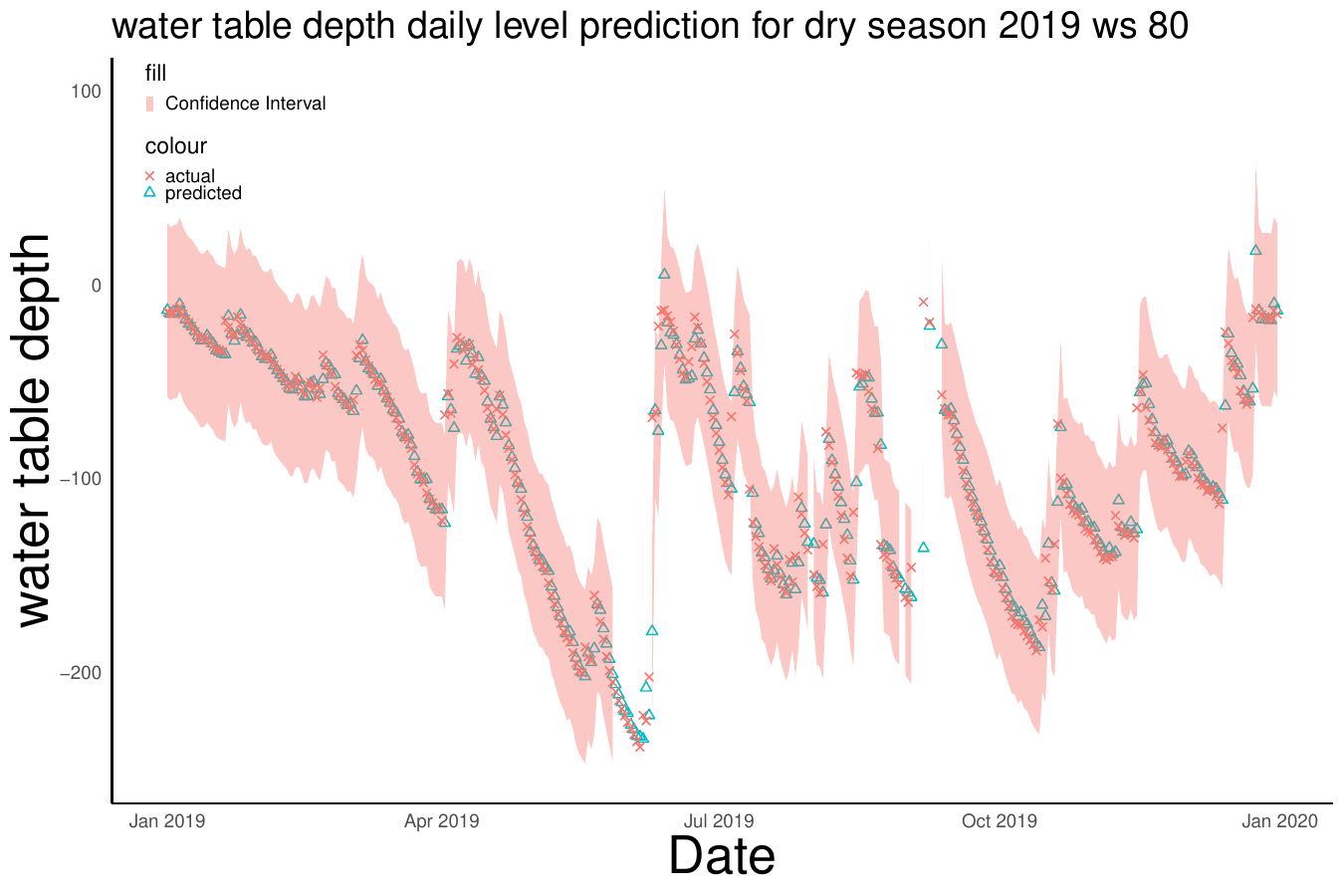} }}%
    \qquad
    \subfloat[\centering ]{{\includegraphics[width=16cm,height=0.4\textheight]{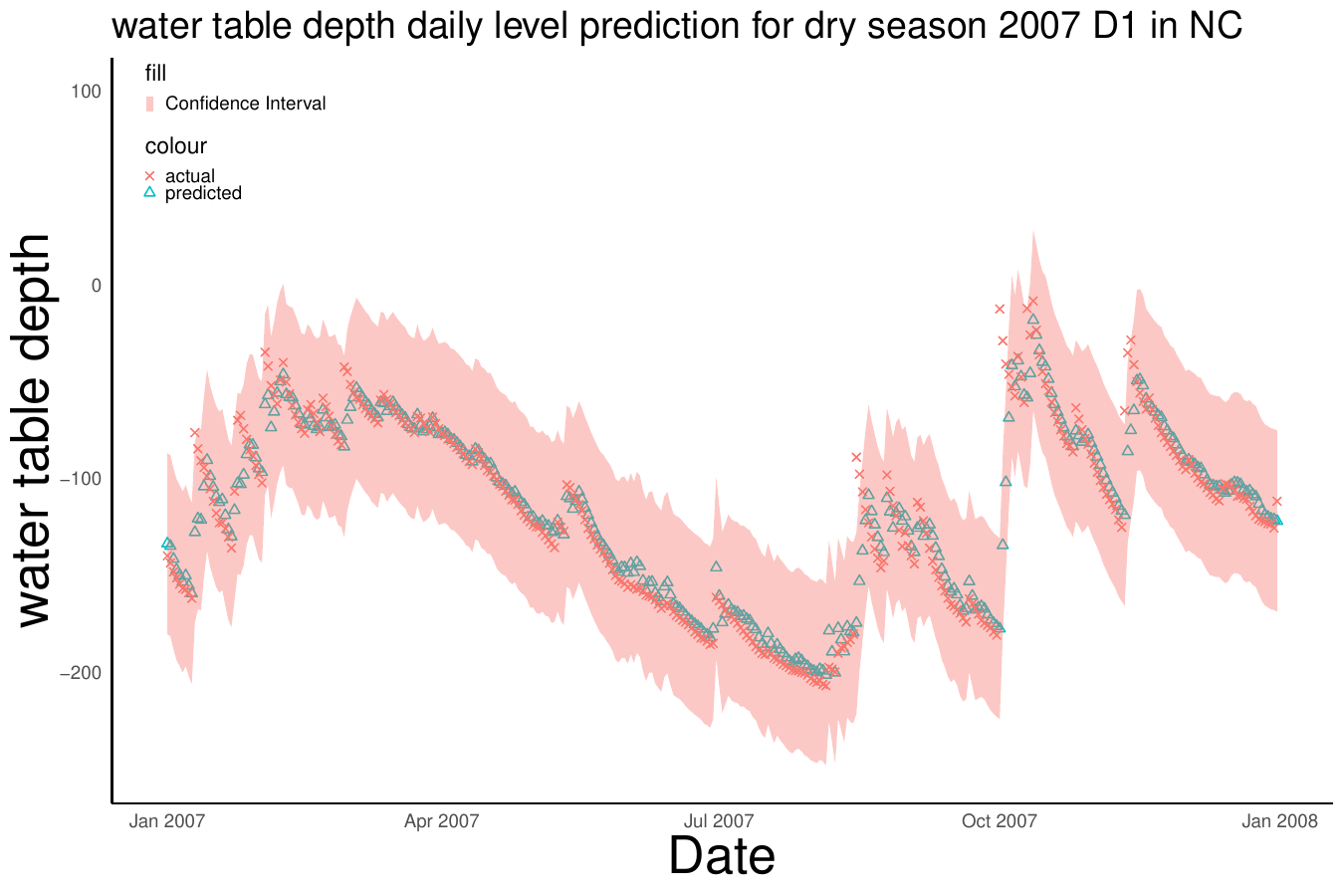} }}%
     \caption{
     Time series plots of predicted vs measured in cm for dry year 2019 WS 80 in SC and 2007 D1 in NC }%
    \label{fig:ts_plots_dry_ws80_2017_NC_2007}%
\end{figure}

\begin{figure}%
    \centering
    \subfloat[\centering]{{\includegraphics[width=16cm,height=.4\textheight]{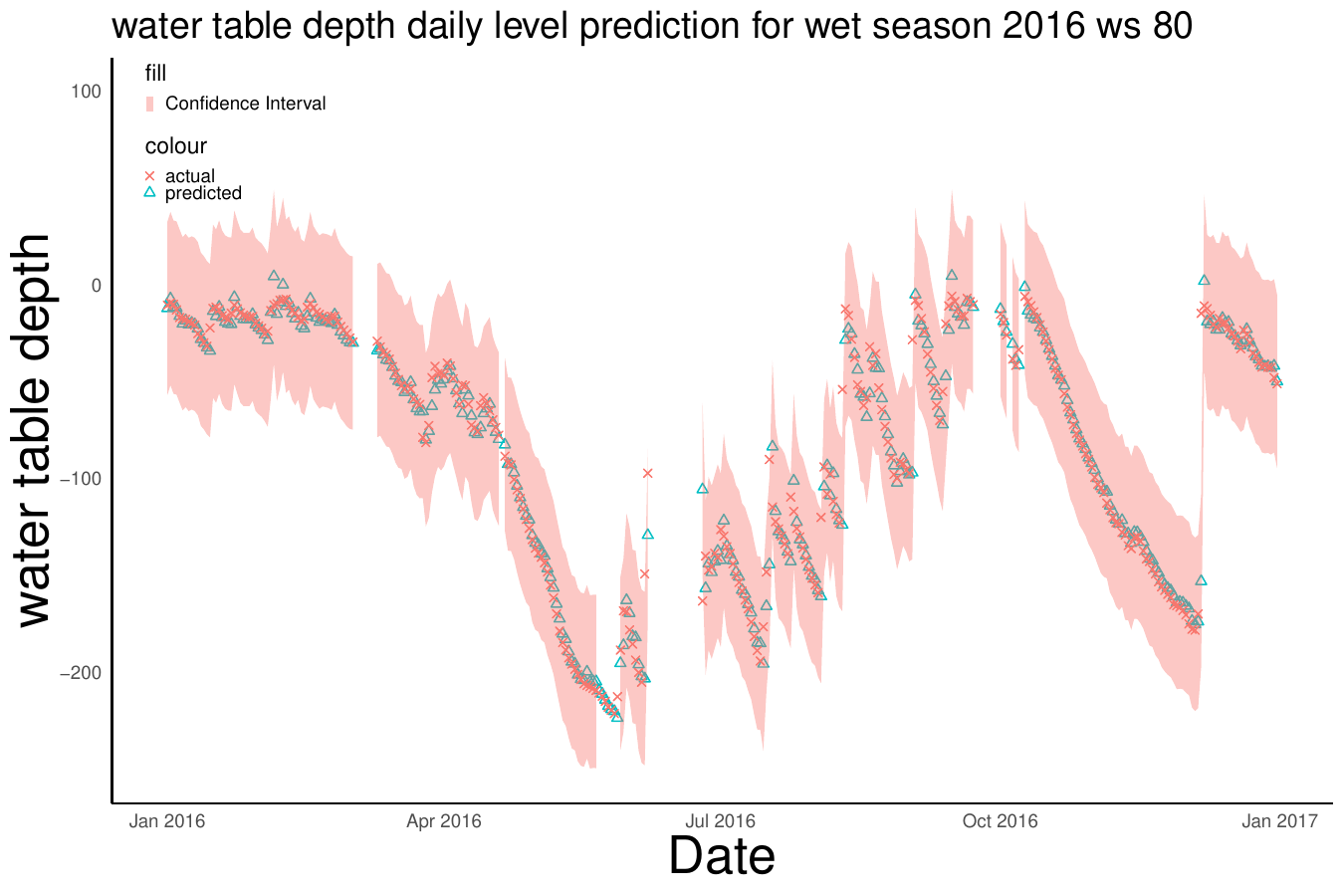} }}%
    \qquad
    \subfloat[\centering]{{\includegraphics[width=16cm,height=0.4\textheight]{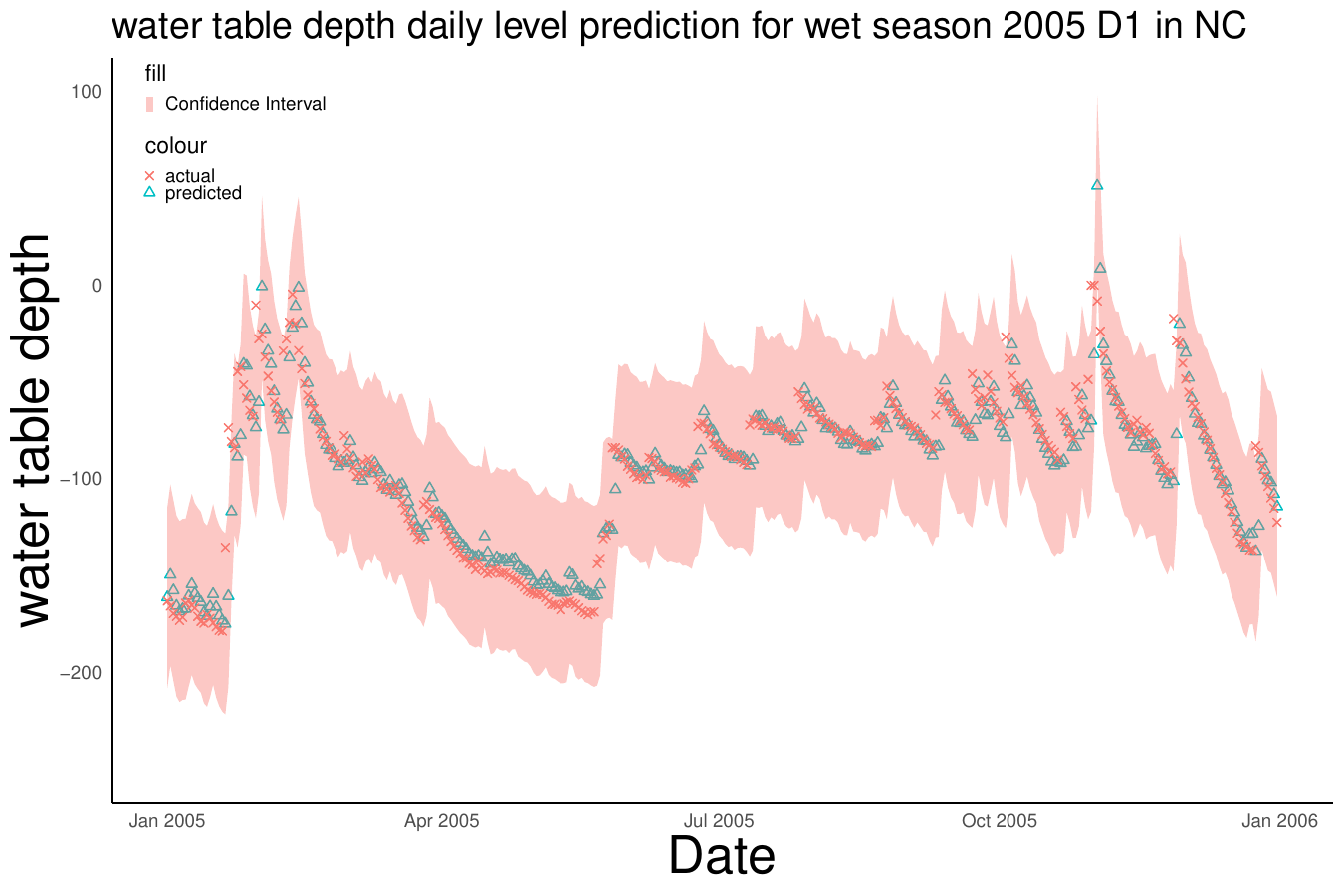} }}%
     \caption{Time series plots of predicted vs measured in cm for wet year 2016 WS 80 in SC and 2005 D1 in NC }%
    \label{fig:ts_plots_wet_ws80_2016_NC_2005}%
\end{figure}

\begin{figure}[h!]%
    \centering
    \subfloat[\centering ]{{\includegraphics[width=16cm,height=.6\textheight]{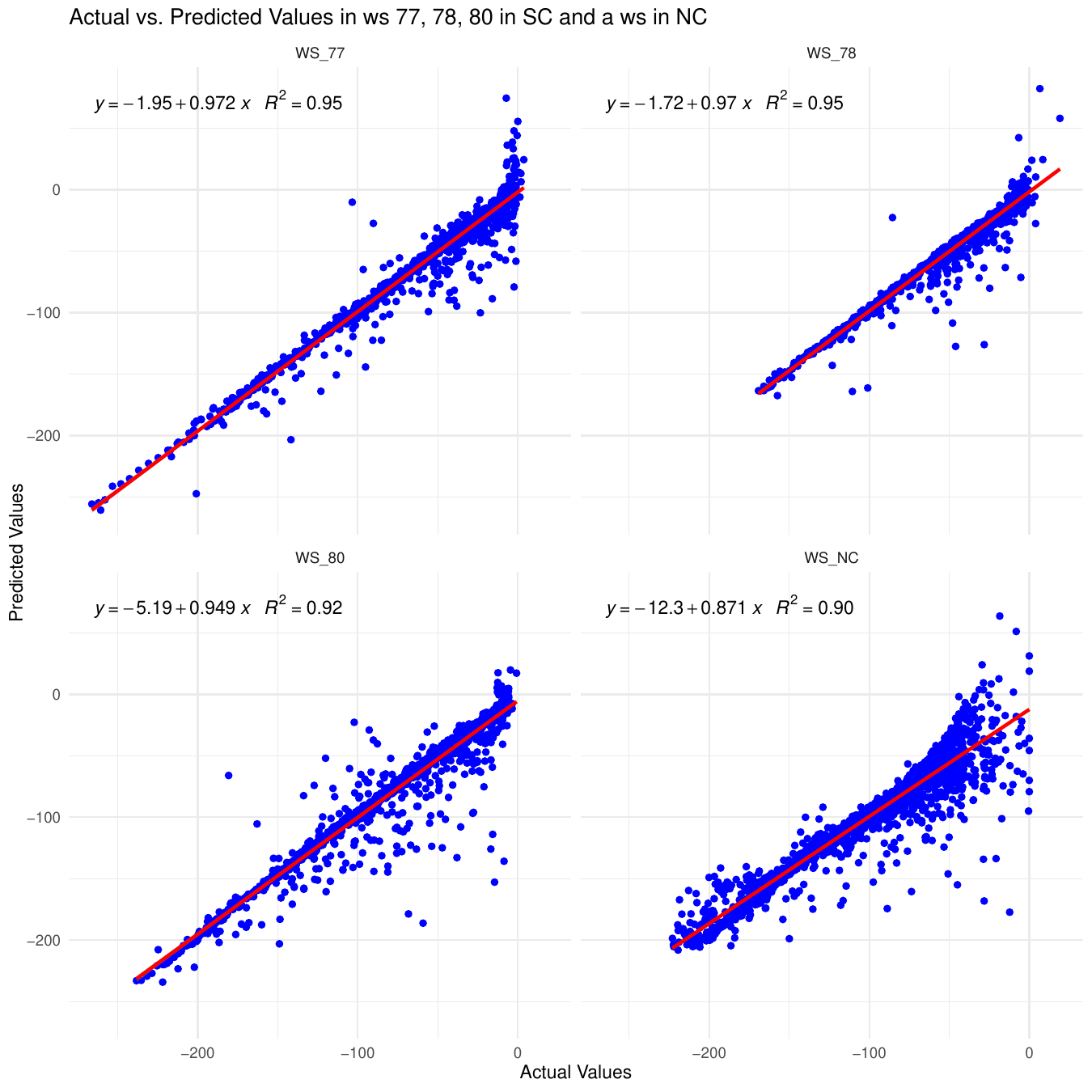} }}%
     \caption{Scatter plots of predicted and measured daily water table depths in cm for watersheds WS 77, WS78, WS80 in SC and D1 in NC
     }%
    \label{fig:regression_plot}%
\end{figure}

In the figures ~\ref{fig:ts_plots_77_78}, and ~\ref{fig:ts_plots_80_NC}, the actual vs prediction of water table depth for 3 different watersheds from South Carolina WS77, WS78, WS80, and one watershed D1 from North Carolina over time series for a daily-level prediction is plotted. The actual measurement is given as a cross $(\times)$ and the predicted measurement is given as a delta $(\Delta)$. In the plot of water table depths of 2017 for WS80 in SC and the year 2007, D1 in NC in figure \ref{fig:ts_plots_dry_ws80_2017_NC_2007} to obtain a clear picture of our prediction for a very dry year and 2016 for WS80 in SC and the year 2005, D1 in NC in the figure \ref{fig:ts_plots_wet_ws80_2016_NC_2005} to obtain a clear picture about our prediction for relatively wet year. The actual versus predicted plot across three different well locations in watersheds WS77 and WS78 in SC and one in watershed D1 in NC is provided in figure \ref{fig:regression_plot}. The total training period is considered 1 to $T_{2}$ time interval and the testing interval is $T_{2}+1$ through $T$ with the horizon as 1. A point estimate gives only a single value as the best estimate of the population parameter. However, a confidence interval provides a range of plausible values for that parameter, giving a better understanding of the uncertainty associated with the estimate (see \cite{manna2024intervalestimationcoefficientspenalized}). A confidence interval for each prediction is included where $\text{SE}=\sqrt{\sum_{t=T_{2}+1}^{T}(y_{t}-\hat{y}_{t})^2}$ as
\[\left(\hat{y}_{t}-3.\text{SE},\hat{y}_{t}+3.\text{SE}\right); t= T_{2}+1,\cdots,T.\]

\begin{table}[H]
\centering
\scriptsize
\begin{threeparttable}
\caption{Residual analysis for water table depth at daily level}
\label{tab:water_table_depth_daily}
\begin{tabular}{lcccccc} 
\toprule
well on Watershed & Reg line (y=) & adj R sq & MSE & RMSE & Nash-Sutcliffe Efficiency \\
\midrule
WS77 & -0.82 + 0.98x & 0.95 & 116.37 & 10.79 & 0.95 \\
WS78 & -1.00 + 0.97x & 0.94 & 75.13 & 8.67 & 0.95 \\
WS80 & -0.86 + 0.97x & 0.92 & 223.28 & 14.94 & 0.92 \\
D1 & -3.05 + 1.03x & 0.90 & 242.39 & 15.57 & 0.90 \\
\bottomrule
\end{tabular}
\begin{tablenotes}
    \small
    \item[a] \scriptsize{In Reg line, y is actual, and x is predicted.}
    \item[b] \scriptsize{MSE is measured in sq cm.}
\end{tablenotes}
\end{threeparttable}
\end{table}

\begin{center}
\scriptsize
\begin{table}[!ht]
\centering
\begin{threeparttable}[H]
\caption{Residual analysis for water table depth for SC monthly, growing and dormant season}
\label{tab:water_table_depth_monthly_seasonal}
\scriptsize
\begin{tabular}{l c c c c c} 
\hline\hline 
Watershed & adj R sq & MSE & RMSE & Nash-Sutcliffe Efficiency  
\\ [0.5ex]
\hline 

WS77 Monthly  &.14 & 922.94 & 30.38 & .14  \\
WS77 Growing &.93 & 164.86  & 12.84 & .93  \\
WS77 Dormant &.93 & 74.88 & 8.65 & .93  \\

\hline 
WS78 Monthly  &.56 & 489.50  & 22.12 & .56  \\
WS78 Growing &.90 & 63.11 & 7.94 & .90  \\
WS78 Dormant &.89 & 50.56 & 7.11 & .89  \\

\hline 
WS80 Monthly  &.16 & 1479.47 & 38.46 & .16  \\
WS80 Growing &.90 & 304.14 & 17.44 & .90  \\
WS80 Dormant &.94 & 101.81 & 10.09 & .94  \\
\hline 
\end{tabular}
\label{tab:monthly_prediction}
\begin{tablenotes}
      \small
      \item \scriptsize {MSE is measured in sq cm.}
    \end{tablenotes}
\end{threeparttable}
\end{table}
\end{center}
\vspace{-1cm}
A table describing the performance of proposed model for daily level in SC WS data is given in table \ref{tab:water_table_depth_daily}. Another table is presented with the performance of proposed model for SC WS data in a few other seasonal variants in table \ref{tab:water_table_depth_monthly_seasonal}. A detailed analysis of goodness of fit measures based on the ``HydroGof'' package (see \cite{zambrano2017package}) for different seasonal variation comparisons are provided in table \ref{tab:water_table_depth_detailed_1} and table \ref{tab:water_table_depth_detailed_2}.
\begin{table}[ht!]
\centering
\begin{threeparttable}
\caption{Detailed Residual analysis for water table depth for SC monthly, growing and dormant season part 1}
\label{tab:water_table_depth_detailed_1}
\scriptsize
\begin{tabular}{rrrrrrr}
  \hline
 & WS77\_daily & WS77\_monthly & WS77\_growing & WS77\_dormant & WS78\_daily & WS78\_monthly \\ 
  \hline
ME & 0.29 & -0.03 & 0.29 & 0.61 & 0.19 & -3.67 \\ 
  MAE & 5.52 & 23.93 & 6.91 & 4.08 & 4.06 & 18.54 \\ 
  MSE & 116.37 & 922.94 & 164.86 & 74.88 & 75.13 & 489.50 \\ 
  RMSE & 10.79 & 30.38 & 12.84 & 8.65 & 8.67 & 22.12 \\ 
  ubRMSE & 10.78 & 30.38 & 12.84 & 8.63 & 8.67 & 21.82 \\ 
  NRMSE \% & 21.80 & 91.60 & 25.60 & 26.40 & 23.20 & 65.40 \\ 
  PBIAS \% & -0.50 & 0.10 & -0.40 & -1.60 & -0.40 & 7.40 \\ 
  RSR & 0.22 & 0.92 & 0.26 & 0.26 & 0.23 & 0.65 \\ 
  rSD & 1.00 & 0.80 & 1.02 & 1.01 & 1.00 & 0.83 \\ 
  NSE & 0.95 & 0.14 & 0.93 & 0.93 & 0.95 & 0.56 \\ 
  mNSE & 0.86 & 0.06 & 0.82 & 0.83 & 0.86 & 0.32 \\ 
  rNSE & -39.43 & -0.81 & -1.14 & -8.68 & 0.54 & -11.02 \\ 
   wNSE & 0.98 & -0.27 & 0.97 & 0.96 & 0.96 & 0.55 \\ 
  d & 0.99 & 0.70 & 0.98 & 0.98 & 0.99 & 0.85 \\ 
  dr & 0.93 & 0.53 & 0.91 & 0.92 & 0.93 & 0.66 \\ 
  md & 0.93 & 0.48 & 0.91 & 0.92 & 0.93 & 0.62 \\ 
  rd & -9.14 & 0.36 & 0.47 & -1.42 & 0.89 & -3.02 \\ 
  cp & 0.51 & 0.42 & 0.48 & 0.38 & 0.36 & 0.47 \\ 
  r & 0.98 & 0.49 & 0.97 & 0.97 & 0.97 & 0.76 \\ 
  R2 & 0.95 & 0.14 & 0.93 & 0.93 & 0.95 & 0.56 \\ 
  bR2 & 0.94 & 0.11 & 0.93 & 0.91 & 0.94 & 0.52 \\ 
  KGE & 0.98 & 0.45 & 0.96 & 0.96 & 0.97 & 0.70 \\ 
  KGElf & -1.77 & 0.02 & -9.56 & 0.37 & 0.19 & -44.23 \\ 
  KGEnp & 0.96 & 0.47 & 0.95 & 0.94 & 0.97 & 0.75 \\ 
  VE & 1.09 & 1.45 & 1.10 & 1.11 & 1.08 & 1.38 \\ 
   \hline
\end{tabular}
\begin{tablenotes}
      \small
      \scriptsize {MSE is measured in sq cm.}
    \end{tablenotes}
\end{threeparttable}
\end{table}

\begin{table}[ht]
\centering
\begin{threeparttable}
\caption{Detailed Residual analysis for water table depth for SC monthly, growing and dormant season part 2}
\label{tab:water_table_depth_detailed_2}
\scriptsize
\begin{tabular}{rrrrrrr}
  \hline
& WS78\_growing & WS78\_dormant & WS80\_daily & WS80\_monthly & WS80\_growing & WS80\_dormant \\ 
  \hline
ME & -1.49 & -0.18 & 1.24 & 7.74 & 1.30 & 1.22 \\ 
  MAE & 4.86 & 3.26 & 6.95 & 31.29 & 8.88 & 4.48 \\ 
  MSE & 63.11 & 50.56 & 223.28 & 1479.47 & 304.14 & 101.81 \\ 
  RMSE & 7.94 & 7.11 & 14.94 & 38.46 & 17.44 & 10.09 \\ 
  ubRMSE & 7.80 & 7.11 & 14.89 & 37.68 & 17.39 & 10.02 \\ 
  NRMSE \% & 31.60 & 32.70 & 28.10 & 90.50 & 31.60 & 24.50 \\ 
  PBIAS \% & 2.30 & 0.60 & -1.60 & -9.20 & -1.40 & -2.20 \\ 
  RSR & 0.32 & 0.33 & 0.28 & 0.91 & 0.32 & 0.24 \\ 
  rSD & 1.04 & 0.99 & 1.01 & 1.00 & 1.03 & 1.01 \\ 
  NSE & 0.90 & 0.89 & 0.92 & 0.16 & 0.90 & 0.94 \\ 
  mNSE & 0.77 & 0.81 & 0.84 & 0.03 & 0.81 & 0.87 \\ 
  rNSE & 0.90 & 0.44 & -2.46 & -0.76 & 0.63 & 0.25 \\ 
  wNSE & 0.89 & 0.94 & 0.92 & -0.10 & 0.90 & 0.93 \\ 
  d & 0.98 & 0.97 & 0.98 & 0.76 & 0.98 & 0.99 \\ 
  dr & 0.88 & 0.91 & 0.92 & 0.52 & 0.91 & 0.93 \\ 
  md & 0.89 & 0.91 & 0.92 & 0.55 & 0.91 & 0.93 \\ 
  rd & 0.98 & 0.86 & 0.14 & 0.51 & 0.91 & 0.81 \\ 
  cp & 0.39 & 0.27 & 0.25 & 0.03 & 0.31 & 0.18 \\ 
  r & 0.95 & 0.95 & 0.96 & 0.60 & 0.95 & 0.97 \\ 
  R2 & 0.90 & 0.89 & 0.92 & 0.16 & 0.90 & 0.94 \\ 
  bR2 & 0.88 & 0.88 & 0.90 & 0.14 & 0.89 & 0.92 \\ 
  KGE & 0.93 & 0.95 & 0.96 & 0.59 & 0.94 & 0.96 \\ 
  KGElf & 0.93 & 0.36 & 0.39 & -0.06 & 0.38 & 0.53 \\ 
  KGEnp & 0.94 & 0.96 & 0.95 & 0.52 & 0.94 & 0.96 \\ 
  VE & 1.08 & 1.11 & 1.09 & 1.37 & 1.10 & 1.08 \\ 
   \hline
\end{tabular}
\begin{tablenotes}
      \small
      \scriptsize{MSE is measured in sq cm.}
\end{tablenotes}
\end{threeparttable}
\end{table}
The following tables provide a detailed analysis of residual metrics for water table depth across different seasonal conditions and years. In the first table (Table \ref{tab:water_table_depth_detailed_1}), coefficients are presented for several performance metrics across multiple periods. For instance, ME (Mean Error) varies notably between different seasonal conditions and years, for example, 0.61 in WS77\_dormant. MAE (Mean Absolute Error) shows considerable differences as well, with WS80\_daily exhibiting the highest value of 8.88, contrasting with the lowest value of 4.06 in WS78\_daily.

Meanwhile, metrics like MSE (Mean Squared Error) and RMSE (Root Mean Squared Error) reflect the variability in model accuracy across different conditions. For example, WS77\_growing shows MSE 164.86 and RMSE of 12.84, whereas WS77\_dormant shows MSE of 74.88 and RMSE of 8.65, indicating better model performance during dormant periods.
The table (table \ref{tab:water_table_depth_monthly_seasonal}) presents a detailed comparison of residual metrics for water table depth across different seasonal conditions (growing, and dormant) in various watersheds (WS77, WS78, WS80). 

For WS77, the model's adjusted R-squared values vary significantly across seasons: growing (0.93), and dormant (0.93). Correspondingly, the Mean Squared Error (MSE) ranges from 164.86 (growing) to 74.88 (dormant), reflecting differing levels of model accuracy. Root Mean Squared Error (RMSE) follows a similar trend, with values ranging from 12.84 (growing) to 8.65 (dormant), indicating higher precision during dormant periods.

In WS78, the adjusted R-squared values are moderately high across all seasons: growing (0.90), and dormant (0.89). The MSE decreases notably from 63.11 (growing) to 50.56 (dormant), suggesting improved model performance during dormant periods. RMSE also decreases accordingly, with values ranging from 7.94 (growing) to 7.11 (dormant).

For WS80, the model's adjusted R-squared values vary widely: growing (0.90), and dormant (0.94). The MSE is highest in the growing period (304.14) and lowest during the dormant season (101.81), indicating varying levels of accuracy. RMSE values range from 17.44 (growing) to 10.09 (dormant), illustrating the model's ability to predict more accurately during dormant conditions.

NRMSE is lower during the dormant period for WS80, indicating better model performance during this phase. The growing season has the highest NRMSE, particularly for WS78 and WS80, indicating more variability in predictions during that period. NSE values are consistently high across all sites and periods with the daily showing slightly higher efficiency values compared to dormant and growing seasons. The weighted NSE, like the standard NSE, shows higher values during the daily period.

These metrics provide insights into the performance and variability of water table depth predictions across different environmental contexts, aiding in better understanding and management of water resources in agricultural and environmental applications.

\subsection{Variable selection}
\label{Variable selection}
Variable selection allows researchers to identify the most relevant variables to include in a model, leading to improved model performance, better interpretability, and reduced overfitting by eliminating unnecessary variables that could introduce noise and complexity to the analysis. In this section, we are demonstrating the important variables which are sufficient for a practitioner to build a daily water table predictive model. In the tables \ref{table:ws_77}, \ref{table:ws_78} and \ref{table:ws_80}, the coefficients of the models for water table depth for 3 different watersheds with 4 different time setups are provided. In this model, elastic net, a penalization over the coefficients approach, is used for obtaining the inference. If a coefficient is closer to zero, the corresponding variable is less sensitive than the other variables away from zero. The table ``Coefficients of WS 77'' presents regression coefficients for various models applied to Watershed 77 (WS77). These models are used to predict a dependent variable based on multiple independent variables. Each column represents a different time scale or season: daily, monthly, growing season, and dormant season. Intercept is the constant term in the regression equation. It represents the expected value of the dependent variable when all predictors are zero. Y1L1, Y1L2, Y1L3, and Y1L4 are lagged values of the dependent variable. These coefficients indicate the influence of previous values of the dependent variable on the current value.
$\text{Air\_Temp\_C1}$ is the coefficient for air temperature at time lag 1 and $\text{Air\_Temp\_C2}$ is the coefficient for air temperature at time lag 2. This notation applies to each of the other variables. This table illustrates how the influence of various factors on the dependent variable differs across daily, monthly, growing season, and dormant season periods. Zero values indicate that the corresponding variable was not included in the model for the specific period.
For the intercepts: WS77\_daily: -0.605, WS77\_growing: -0.818, WS77\_dormant: 3.842
Among the predictors, notable differences across conditions are observed:
Y1L1 has consistent positive coefficients across all conditions, ranging approximately from 0.573 to 0.945.
Rainfall1 shows variations, with coefficients of approximately 0.459 for WS77\_daily, 0.505 for WS77\_growing and around 0.462 for WS77\_dormant.
DailyFlow1 has negative coefficients, ranging from $-0.032$ to $-0.478$, with WS77\_growing exhibiting the most negative impact. Wind direction with 2nd order lag might have some small influence except the monthly prediction. Certain predictors such as $\text{Air\_Temp\_C1}$, $\text{Soil\_Temp\_1}$, $\text{net\_rad1}$, $\text{PET1}$, and $\text{RH1}$ have coefficients of zero across all conditions, indicating no significant impact in these scenarios. Similar patterns are observed in other predictors across different conditions, reflecting their varying degrees of influence.
In summary, the coefficients vary widely across the different conditions of WS77\_daily, WS77\_growing, and WS77\_dormant, underscoring the nuanced relationships between predictors and outcomes in each specific context. These variations highlight the importance of considering environmental and growth conditions when interpreting the effects of predictors on the response variable. For the other 2 watersheds in SC, WS77 and WS78 a similar pattern of important coefficients are observable.

\begin{table}[ht!]
\caption{Coefficients of WS77 }
\label{table:ws_77}
\centering
\scriptsize
\begin{tabular}{rrrrr}
  \hline
 & WS77\_daily & WS77\_monthly & WS77\_growing & WS77\_dormant \\ 
  \hline
intercept & -0.60460214 & -44.17051465 & -0.81780409 & 3.84158477 \\ 
  Y1L1 & 0.94170219 & 0.57386687 & 0.92943288 & 0.94451563 \\ 
  Y1L2 & 0.00000000 & -0.06360627 & 0.00000000 & 0.02414715 \\ 
  Y1L3 & 0.00000000 & 0.06150226 & 0.00000000 & 0.00000000 \\ 
  Y1L4 & 0.02612793 & -0.07874109 & 0.02473702 & 0.01099902 \\ 
  Air\_Temp\_C1 & 0.00000000 & 0.22590743 & 0.00000000 & 0.00000000 \\ 
  Soil\_Temp\_1 & 0.00000000 & 0.00000000 & 0.00000000 & 0.00000000 \\ 
  Rainfall1 & 0.45885660 & 0.00000000 & 0.50514104 & 0.46214280 \\ 
  DailyFlow1 & -0.10814269 & -0.03207597 & -0.47766625 & -0.04136039 \\ 
  Solar\_rad\_1 & -0.03770686 & 0.00000000 & -0.03914593 & -0.03038076 \\ 
  net\_rad1 & 0.00000000 & 0.00000000 & 0.00000000 & 0.00000000 \\ 
  PET1 & 0.00000000 & 0.00000000 & 0.00000000 & 0.00000000 \\ 
  RH1 & 0.00000000 & 0.00000000 & 0.00000000 & 0.00000000 \\ 
  Windspeed\_1 & 0.00000000 & -0.04562469 & 0.00000000 & 0.00000000 \\ 
  Wind\_direc1 & -0.00119424 & 0.00000000 & 0.00885937 & -0.01274562 \\ 
  Vapor\_Kpa1 & 0.00000000 & 0.25763440 & 0.00000000 & 0.00000000 \\ 
  Air\_Temp\_C2 & 0.00000000 & 0.00000000 & 0.00000000 & 0.00000000 \\ 
  Soil\_Temp\_2 & 0.00000000 & -0.99438617 & 0.00000000 & 0.00000000 \\ 
  Rainfall2 & 0.00000000 & -0.80569544 & 0.00957783 & 0.00000000 \\ 
  DailyFlow2 & 0.00000000 & -0.28049803 & 0.03664337 & 0.01046831 \\ 
  Solar\_rad\_2 & 0.01681527 & 0.00000000 & 0.01030524 & 0.00143884 \\ 
  net\_rad2 & 0.00000000 & 0.00000000 & 0.00000000 & 0.00000000 \\ 
  PET2 & 0.00000000 & 0.00000000 & 0.00000000 & 0.00000000 \\ 
  RH2 & 0.00000000 & 0.00000000 & 0.00000000 & 0.00000000 \\ 
  Windspeed\_2 & 0.00000000 & 0.29262960 & 0.00000000 & 0.00000000 \\ 
  Wind\_direc2 & 0.00363181 & 0.00000000 & 0.00063054 & -0.00263301 \\ 
  Vapor\_Kpa2 & 0.00000000 & 0.16987948 & 0.00000000 & 0.00000000 \\ 
   \hline
\end{tabular}
\end{table}

\begin{table}[ht!]
\caption{Coefficients of WS78}
\label{table:ws_78}
\centering
\scriptsize
\begin{tabular}{rrrrr}
  \hline
 & WS78\_daily & WS78\_monthly & WS78\_growing & WS78\_dormant \\ 
  \hline
intercept & 1.04016246 & -17.78534371 & 2.02273146 & 2.25925001 \\ 
  Y1L1 & 0.96681749 & 0.48754284 & 0.94329889 & 0.95284414 \\ 
  Y1L2 & 0.00000000 & 0.02230166 & 0.00000000 & 0.01951081 \\ 
  Y1L3 & 0.00000000 & 0.05422990 & 0.00000000 & 0.00000000 \\ 
  Y1L4 & 0.00000000 & 0.00000000 & 0.00000000 & 0.00000000 \\ 
  Air\_Temp\_C1 & 0.00000000 & 0.00000000 & 0.00000000 & 0.00000000 \\ 
  Soil\_Temp\_1 & 0.00000000 & 0.55704770 & 0.00000000 & 0.00000000 \\ 
  Rainfall1 & 0.21339954 & -2.26025056 & 0.17364045 & 0.19415649 \\ 
  DailyFlow1 & -0.05956410 & 0.01723743 & -0.01778909 & 0.00000000 \\ 
  Solar\_rad\_1 & -0.03263608 & 0.00000000 & -0.04168843 & -0.02497973 \\ 
  net\_rad1 & 0.00000000 & 0.32158411 & 0.00000000 & 0.00000000 \\ 
  PET1 & 0.00000000 & 0.00000000 & 0.00000000 & 0.00000000 \\ 
  RH1 & 0.00000000 & 0.00000000 & 0.00000000 & 0.00000000 \\ 
  Windspeed\_1 & 0.00000000 & 0.00375495 & 0.00000000 & 0.00000000 \\ 
  Wind\_direc1 & 0.00000000 & 0.00000000 & 0.00736859 & -0.00842944 \\ 
  Vapor\_Kpa1 & 0.00000000 & 0.23684159 & 0.00000000 & 0.00000000 \\ 
  Air\_Temp\_C2 & 0.00000000 & -1.29848452 & 0.00000000 & 0.00000000 \\ 
  Soil\_Temp\_2 & 0.00000000 & -0.11295115 & 0.00000000 & 0.00000000 \\ 
  Rainfall2 & -0.01749112 & -3.24053373 & -0.01778637 & 0.00000000 \\ 
  DailyFlow2 & 0.00000000 & -0.40208839 & 0.00000000 & 0.00000000 \\ 
  Solar\_rad\_2 & 0.00965442 & -0.03255979 & 0.00483663 & 0.00208397 \\ 
  net\_rad2 & 0.00000000 & 0.00000000 & 0.00000000 & 0.00000000 \\ 
  PET2 & 0.00000000 & 0.00000000 & 0.00000000 & 0.00000000 \\ 
  RH2 & 0.00000000 & 0.00000000 & 0.00000000 & 0.00000000 \\ 
  Windspeed\_2 & 0.00000000 & 0.18638101 & 0.00000000 & 0.00000000 \\ 
  Wind\_direc2 & 0.00167598 & 0.00000000 & 0.00000000 & 0.00000000 \\ 
  Vapor\_Kpa2 & 0.00000000 & 0.20761335 & 0.00000000 & 0.00000000 \\ 
   \hline
\end{tabular}
\end{table}

\begin{table}[ht!]
\caption{coefficients of WS80}
\label{table:ws_80}
\centering
\scriptsize
\begin{tabular}{rrrrr}
  \hline
 & WS80\_daily & WS80\_monthly & WS80\_growing & WS80\_dormant \\ 
  \hline
intercept & 0.52509421 & 13.18754808 & -0.13058237 & 2.92482323 \\ 
  Y1L1 & 0.93421150 & 0.58294513 & 0.93060155 & 0.93285971 \\ 
  Y1L2 & 0.00000000 & 0.00000000 & 0.00000000 & 0.00000000 \\ 
  Y1L3 & 0.00000000 & 0.08972491 & 0.00000000 & 0.00000000 \\ 
  Y1L4 & 0.03535136 & -0.02303886 & 0.02326997 & 0.04122628 \\ 
  Air\_Temp\_C1 & 0.00000000 & 0.00000000 & 0.00000000 & 0.00000000 \\ 
  Soil\_Temp\_1 & 0.00000000 & 0.00000000 & 0.00000000 & 0.00000000 \\ 
  Rainfall1 & 0.35935032 & 0.00000000 & 0.39804321 & 0.26960723 \\ 
  DailyFlow1 & 0.00000000 & 0.00000000 & 0.00000000 & 0.00000000 \\ 
  Solar\_rad\_1 & -0.01813562 & -0.08842600 & -0.00861104 & -0.01779213 \\ 
  net\_rad1 & -0.01601925 & 0.00000000 & -0.04803728 & 0.00000000 \\ 
  PET1 & 0.00000000 & 0.00000000 & 0.00000000 & 0.00000000 \\ 
  RH1 & 0.00000000 & 0.00000000 & 0.00000000 & 0.00000000 \\ 
  Windspeed\_1 & 0.00000000 & -0.02571220 & 0.00000000 & 0.00000000 \\ 
  Wind\_direc1 & -0.00266504 & 0.00000000 & 0.00000000 & -0.01055811 \\ 
  Vapor\_Kpa1 & 0.00000000 & 0.35877366 & 0.00000000 & 0.00000000 \\ 
  Air\_Temp\_C2 & 0.00000000 & 0.00000000 & 0.00000000 & 0.00000000 \\ 
  Soil\_Temp\_2 & 0.00000000 & 0.00000000 & 0.00000000 & 0.00000000 \\ 
  Rainfall2 & 0.00000000 & 0.00000000 & 0.00000000 & 0.00000000 \\ 
  DailyFlow2 & 0.00000000 & -0.20660409 & 0.00000000 & 0.00000000 \\ 
  Solar\_rad\_2 & 0.00010972 & -0.40075777 & 0.00224428 & -0.00712781 \\ 
  net\_rad2 & 0.00000000 & 0.00000000 & 0.00000000 & -0.01884551 \\ 
  PET2 & 0.00000000 & 0.00000000 & 0.00000000 & 0.00000000 \\ 
  RH2 & 0.00000000 & 0.00000000 & 0.00000000 & 0.00000000 \\ 
  Windspeed\_2 & 0.00000000 & -0.06453808 & 0.00000000 & 0.00000000 \\ 
  Wind\_direc2 & 0.00015576 & 0.00000000 & 0.00027031 & 0.00000000 \\ 
  Vapor\_Kpa2 & 0.00000000 & 0.11455799 & 0.00000000 & 0.00000000 \\ 
   \hline
\end{tabular}
\end{table}

\subsection{Exploration without daily-flow variable}
In the section \ref{Variable selection}, we observed that daily flow with it's one and two day's lag is important for WS 77 and WS 78 water table depth prediction. The investigation how the prediction might have changed if the daily flow variable is absent is also important as daily flow data is not frequently available like other climatic parameters. The coefficients are described in table \ref{table:ws_77_no_dailyflow}, \ref{table:ws_78_no_dailyflow} and \ref{table:ws_78_no_dailyflow} for 3 different stations in SC. The residuals are listed in table \ref{tab:water_table_depth_detailed_nodailyflow_1} and \ref{tab:water_table_depth_detailed_nodailyflow_2}. The set of important coefficients did not change significantly. In terms of daily level prediction, the mean squared error (MSE) for WS77 changed from 116.37 to 116.81, for WS78 changed from 75.13 to 75.68, and for WS80 almost no change was observed. For monthly prediction, the MSE for WS77 changed from 922.94 to 936.28, for WS78 changed from 489.50 to 326.32, and for WS80 almost no change was observed. During the growing period, the MSE for WS77 changed from 164.86 to 201.49, for WS78 changed from 63.11 to 63.30, and for WS80 almost no change was observed. During the dormant period, the MSE for WS77 changed from 74.88 to 82.44, and for WS78 and WS80, almost no change was observed. Hence, dropping the daily-flow variable did not result in a significant change in the model prediction.
\begin{table}[ht!]
\caption{Coefficients of WS77 without daily-flow variable}
\label{table:ws_77_no_dailyflow}
\centering
\scriptsize
\begin{tabular}{rrrrr}
  \hline
 & WS77\_daily & WS77\_monthly & WS77\_growing & WS77\_dormant \\ 
  \hline
intercept & -0.85632733 & -44.72741826 & 0.09830566 & 2.79063069 \\ 
  Y1L1 & 0.93853701 & 0.57162827 & 0.97742642 & 0.94503564 \\ 
  Y1L2 & 0.00000000 & -0.06954435 & -0.03778470 & 0.01449843 \\ 
  Y1L3 & 0.00000000 & 0.06535313 & 0.00000000 & 0.00000000 \\ 
  Y1L4 & 0.02807640 & -0.07969383 & 0.01013189 & 0.01495053 \\ 
  Air\_Temp\_C1 & 0.00000000 & 0.24341318 & 0.00000000 & 0.00000000 \\ 
  Soil\_Temp\_1 & 0.00000000 & 0.00000000 & 0.00000000 & 0.00000000 \\ 
  Rainfall1 & 0.45628644 & -0.03634609 & 0.30991193 & 0.37887521 \\ 
  Solar\_rad\_1 & -0.03762190 & 0.00000000 & -0.04888764 & -0.03384639 \\ 
  net\_rad1 & 0.00000000 & 0.00000000 & 0.00000000 & 0.00000000 \\ 
  PET1 & 0.00000000 & 0.00000000 & 0.00000000 & 0.00000000 \\ 
  RH1 & 0.00000000 & 0.00000000 & 0.00000000 & 0.00000000 \\ 
  Windspeed\_1 & 0.00000000 & -0.04789414 & 0.00000000 & 0.00000000 \\ 
  Wind\_direc1 & -0.00154098 & 0.00000000 & 0.01031562 & -0.01391166 \\ 
  Vapor\_Kpa1 & 0.00000000 & 0.25786024 & 0.00000000 & 0.00000000 \\ 
  Air\_Temp\_C2 & 0.00000000 & 0.00000000 & 0.00000000 & 0.00000000 \\ 
  Soil\_Temp\_2 & 0.00000000 & -1.00013527 & 0.00000000 & 0.00000000 \\ 
  Rainfall2 & 0.00000000 & -0.27353078 & -0.07273526 & 0.00000000 \\ 
  Solar\_rad\_2 & 0.01727760 & 0.00000000 & 0.01592106 & 0.00830007 \\ 
  net\_rad2 & 0.00000000 & 0.00000000 & 0.00000000 & 0.00000000 \\ 
  PET2 & 0.00000000 & 0.00000000 & 0.00000000 & 0.00000000 \\ 
  RH2 & 0.00000000 & 0.00000000 & 0.00000000 & 0.00000000 \\ 
  Windspeed\_2 & 0.00000000 & 0.29280204 & 0.00000000 & 0.00000000 \\ 
  Wind\_direc2 & 0.00393818 & 0.00000000 & 0.00257751 & -0.00117845 \\ 
  Vapor\_Kpa2 & 0.00000000 & 0.16120519 & 0.00000000 & 0.00000000 \\ 
   \hline
\end{tabular}
\end{table}

\begin{table}[ht!]
\caption{Coefficients of WS78 without daily-flow variable}
\label{table:ws_78_no_dailyflow}
\centering
\scriptsize
\begin{tabular}{rrrrr}
  \hline
intercept & 1.03224001 & -27.87703538 & 2.04770545 & 2.25970872 \\ 
  Y1L1 & 0.96589537 & 0.48282910 & 0.94302953 & 0.95285574 \\ 
  Y1L2 & 0.00000000 & 0.01259210 & 0.00000000 & 0.01949983 \\ 
  Y1L3 & 0.00000000 & 0.03278702 & 0.00000000 & 0.00000000 \\ 
  Y1L4 & 0.00000000 & 0.00000000 & 0.00000000 & 0.00000000 \\ 
  Air\_Temp\_C1 & 0.00000000 & 0.00000000 & 0.00000000 & 0.00000000 \\ 
  Soil\_Temp\_1 & 0.00000000 & 0.00000000 & 0.00000000 & 0.00000000 \\ 
  Rainfall1 & 0.20735153 & 0.03302421 & 0.17069105 & 0.19409831 \\ 
  Solar\_rad\_1 & -0.03259846 & 0.00000000 & -0.04181391 & -0.02498287 \\ 
  net\_rad1 & 0.00000000 & 0.00000000 & 0.00000000 & 0.00000000 \\ 
  PET1 & 0.00000000 & 0.00000000 & 0.00000000 & 0.00000000 \\ 
  RH1 & 0.00000000 & 0.00000000 & 0.00000000 & 0.00000000 \\ 
  Windspeed\_1 & 0.00000000 & 0.00000000 & 0.00000000 & 0.00000000 \\ 
  Wind\_direc1 & 0.00000000 & 0.00000000 & 0.00754347 & -0.00842898 \\ 
  Vapor\_Kpa1 & 0.00000000 & 0.16489046 & 0.00000000 & 0.00000000 \\ 
  Air\_Temp\_C2 & 0.00000000 & 0.00000000 & 0.00000000 & 0.00000000 \\ 
  Soil\_Temp\_2 & 0.00000000 & 0.00000000 & 0.00000000 & 0.00000000 \\ 
  Rainfall2 & -0.03251562 & -0.38246738 & -0.02287892 & 0.00000000 \\ 
  Solar\_rad\_2 & 0.00934094 & 0.00000000 & 0.00471329 & 0.00208454 \\ 
  net\_rad2 & 0.00000000 & 0.00000000 & 0.00000000 & 0.00000000 \\ 
  PET2 & 0.00000000 & 0.00000000 & 0.00000000 & 0.00000000 \\ 
  RH2 & 0.00000000 & 0.00000000 & 0.00000000 & 0.00000000 \\ 
  Windspeed\_2 & 0.00000000 & 0.20131652 & 0.00000000 & 0.00000000 \\ 
  Wind\_direc2 & 0.00176242 & 0.00000000 & 0.00000000 & 0.00000000 \\ 
  Vapor\_Kpa2 & 0.00000000 & 0.09310690 & 0.00000000 & 0.00000000 \\ 
   \hline
\end{tabular}
\end{table}

\begin{table}[ht!]
\caption{coefficients of WS80 without daily-flow variable}
\label{table:ws_80_no_dailyflow}
\centering
\scriptsize
\begin{tabular}{rrrrr}
  \hline
 & WS80\_daily & WS80\_monthly & WS80\_growing & WS80\_dormant \\ 
  \hline
intercept & 0.52509101 & 13.18755121 & -0.13058727 & 2.92482270 \\ 
  Y1L1 & 0.93421149 & 0.58294512 & 0.93060154 & 0.93285971 \\ 
  Y1L2 & 0.00000000 & 0.00000000 & 0.00000000 & 0.00000000 \\ 
  Y1L3 & 0.00000000 & 0.08972490 & 0.00000000 & 0.00000000 \\ 
  Y1L4 & 0.03535138 & -0.02303886 & 0.02326999 & 0.04122627 \\ 
  Air\_Temp\_C1 & 0.00000000 & 0.00000000 & 0.00000000 & 0.00000000 \\ 
  Soil\_Temp\_1 & 0.00000000 & 0.00000000 & 0.00000000 & 0.00000000 \\ 
  Rainfall1 & 0.35935068 & 0.00000000 & 0.39804348 & 0.26960720 \\ 
  Solar\_rad\_1 & -0.01813560 & -0.08842595 & -0.00861103 & -0.01779214 \\ 
  net\_rad1 & -0.01601926 & 0.00000000 & -0.04803726 & 0.00000000 \\ 
  PET1 & 0.00000000 & 0.00000000 & 0.00000000 & 0.00000000 \\ 
  RH1 & 0.00000000 & 0.00000000 & 0.00000000 & 0.00000000 \\ 
  Windspeed\_1 & 0.00000000 & -0.02571222 & 0.00000000 & 0.00000000 \\ 
  Wind\_direc1 & -0.00266504 & 0.00000000 & 0.00000000 & -0.01055811 \\ 
  Vapor\_Kpa1 & 0.00000000 & 0.35877367 & 0.00000000 & 0.00000000 \\ 
  Air\_Temp\_C2 & 0.00000000 & 0.00000000 & 0.00000000 & 0.00000000 \\ 
  Soil\_Temp\_2 & 0.00000000 & 0.00000000 & 0.00000000 & 0.00000000 \\ 
  Rainfall2 & 0.00000000 & -0.20660409 & 0.00000000 & 0.00000000 \\ 
  Solar\_rad\_2 & 0.00010971 & -0.40075783 & 0.00224427 & -0.00712774 \\ 
  net\_rad2 & 0.00000000 & 0.00000000 & 0.00000000 & -0.01884564 \\ 
  PET2 & 0.00000000 & 0.00000000 & 0.00000000 & 0.00000000 \\ 
  RH2 & 0.00000000 & 0.00000000 & 0.00000000 & 0.00000000 \\ 
  Windspeed\_2 & 0.00000000 & -0.06453809 & 0.00000000 & 0.00000000 \\ 
  Wind\_direc2 & 0.00015576 & 0.00000000 & 0.00027031 & 0.00000000 \\ 
  Vapor\_Kpa2 & 0.00000000 & 0.11455799 & 0.00000000 & 0.00000000 \\ 
   \hline
\end{tabular}
\end{table}

\begin{table}[ht!]
\caption{Detailed Residual analysis for water table depth for SC monthly, growing and dormant season without daily-flow variable part 1}
\label{tab:water_table_depth_detailed_nodailyflow_1}
\centering
\begin{threeparttable}
\scriptsize
\begin{tabular}{rrrrrrr}
  \hline
 & WS77\_daily & WS77\_monthly & WS77\_growing & WS77\_dormant & WS78\_daily & WS78\_monthly \\ 
  \hline
ME & 0.29 & 0.02 & 0.17 & 0.55 & 0.18 & -2.56 \\ 
  MAE & 5.52 & 24.17 & 7.37 & 4.18 & 4.07 & 15.04 \\ 
  MSE & 116.81 & 936.38 & 201.49 & 82.44 & 75.68 & 326.32 \\ 
  RMSE & 10.81 & 30.60 & 14.19 & 9.08 & 8.70 & 18.06 \\ 
  ubRMSE & 10.80 & 30.60 & 14.19 & 9.06 & 8.70 & 17.88 \\ 
  NRMSE \% & 21.90 & 92.00 & 28.10 & 27.40 & 23.20 & 59.30 \\ 
  PBIAS \% & -0.50 & -0.00 & -0.30 & -1.50 & -0.30 & 5.10 \\ 
  RSR & 0.22 & 0.92 & 0.28 & 0.27 & 0.23 & 0.59 \\ 
  rSD & 1.00 & 0.80 & 1.01 & 1.00 & 1.00 & 0.93 \\ 
  NSE & 0.95 & 0.13 & 0.92 & 0.92 & 0.95 & 0.64 \\ 
  mNSE & 0.86 & 0.05 & 0.81 & 0.83 & 0.86 & 0.41 \\ 
  rNSE & -86.44 & -0.86 & 0.53 & -0.66 & 0.55 & -1.36 \\ 
  wNSE & 0.98 & -0.28 & 0.98 & 0.96 & 0.96 & 0.60 \\ 
  d & 0.99 & 0.69 & 0.98 & 0.98 & 0.99 & 0.89 \\ 
  dr & 0.93 & 0.52 & 0.90 & 0.92 & 0.93 & 0.70 \\ 
  md & 0.93 & 0.48 & 0.90 & 0.92 & 0.93 & 0.68 \\ 
  rd & -20.94 & 0.34 & 0.88 & 0.58 & 0.89 & 0.30 \\ 
  cp & 0.51 & 0.41 & 0.43 & 0.36 & 0.35 & 0.54 \\ 
  r & 0.98 & 0.48 & 0.96 & 0.96 & 0.97 & 0.81 \\ 
  R2 & 0.95 & 0.13 & 0.92 & 0.92 & 0.95 & 0.64 \\ 
  bR2 & 0.94 & 0.11 & 0.91 & 0.90 & 0.94 & 0.62 \\ 
  KGE & 0.98 & 0.44 & 0.96 & 0.96 & 0.97 & 0.79 \\ 
  KGElf & -1.14 & 0.02 & -23.35 & 0.47 & 0.39 & 0.28 \\ 
  KGEnp & 0.96 & 0.45 & 0.95 & 0.94 & 0.97 & 0.82 \\ 
  VE & 1.09 & 1.46 & 1.11 & 1.11 & 1.08 & 1.30 \\ 
   \hline
\end{tabular}
    \begin{tablenotes}
      \small
      \item \scriptsize {MSE is measured in sq cm.}
    \end{tablenotes}
\end{threeparttable}
\end{table}


\begin{table}[ht!]
\caption{Detailed Residual analysis for water table depth for SC monthly, growing and dormant season without daily-flow variable part 2}
\label{tab:water_table_depth_detailed_nodailyflow_2}
\centering
\begin{threeparttable}
\scriptsize
\begin{tabular}{rrrrrrr}
  \hline
& WS78\_growing & WS78\_dormant & WS80\_daily & WS80\_monthly & WS80\_growing & WS80\_dormant \\ 
  \hline
ME & -1.50 & -0.18 & 1.24 & 7.74 & 1.30 & 1.22 \\ 
  MAE & 4.89 & 3.26 & 6.95 & 31.29 & 8.88 & 4.48 \\ 
  MSE & 63.30 & 50.56 & 223.28 & 1479.47 & 304.14 & 101.81 \\ 
  RMSE & 7.96 & 7.11 & 14.94 & 38.46 & 17.44 & 10.09 \\ 
  ubRMSE & 7.81 & 7.11 & 14.89 & 37.68 & 17.39 & 10.02 \\ 
  NRMSE \% & 31.70 & 32.70 & 28.10 & 90.50 & 31.60 & 24.50 \\ 
  PBIAS \% & 2.30 & 0.60 & -1.60 & -9.20 & -1.40 & -2.20 \\ 
  RSR & 0.32 & 0.33 & 0.28 & 0.91 & 0.32 & 0.24 \\ 
  rSD & 1.04 & 0.99 & 1.01 & 1.00 & 1.03 & 1.01 \\ 
  NSE & 0.90 & 0.89 & 0.92 & 0.16 & 0.90 & 0.94 \\ 
  mNSE & 0.77 & 0.81 & 0.84 & 0.03 & 0.81 & 0.87 \\ 
  rNSE & 0.90 & 0.44 & -2.46 & -0.76 & 0.63 & 0.25 \\ 
  wNSE & 0.89 & 0.94 & 0.92 & -0.10 & 0.90 & 0.93 \\ 
  d & 0.98 & 0.97 & 0.98 & 0.76 & 0.98 & 0.99 \\ 
  dr & 0.88 & 0.91 & 0.92 & 0.52 & 0.91 & 0.93 \\ 
  md & 0.89 & 0.91 & 0.92 & 0.55 & 0.91 & 0.93 \\ 
  rd & 0.98 & 0.86 & 0.14 & 0.51 & 0.91 & 0.81 \\ 
  cp & 0.39 & 0.27 & 0.25 & 0.03 & 0.31 & 0.18 \\ 
  r & 0.95 & 0.95 & 0.96 & 0.60 & 0.95 & 0.97 \\ 
  R2 & 0.90 & 0.89 & 0.92 & 0.16 & 0.90 & 0.94 \\ 
  bR2 & 0.88 & 0.88 & 0.90 & 0.14 & 0.89 & 0.92 \\ 
  KGE & 0.93 & 0.95 & 0.96 & 0.59 & 0.94 & 0.96 \\ 
  KGElf & 0.93 & 0.36 & 0.39 & -0.06 & 0.38 & 0.53 \\ 
  KGEnp & 0.94 & 0.96 & 0.95 & 0.52 & 0.94 & 0.96 \\ 
  VE & 1.08 & 1.11 & 1.09 & 1.37 & 1.10 & 1.08 \\ 
   \hline
\end{tabular}
    \begin{tablenotes}
      \small
      \item \scriptsize {MSE is measured in sq cm.}
\end{tablenotes}
\end{threeparttable}
\end{table}

\section{Conclusions and discussion} 
\label{conclusions_discussion}
The vector autoregression (VAR) and vector autoregression with several covariates (VARX)
have served as essential tools in forecasting multivariate time series. In the presence of a high number of covariates, it becomes essential to choose the important covariates. We found that some packages do handle multivariate data with sparsity assumptions such as ``glmnet'', but do not incorporate a time-dependent framework. Also, there is significant research on the autoregressive process with the presence of covariates (VARX), but incorporating variables with their lag is not too much explored. In our scenario, this factor is crucial for example, rainfall of the current day might not immediately affect the current water table depth. Moreover, the water table depth of the previous couple of days might be one key variable that can be used to predict the current water table depth as they are highly correlated. On top of that, we want to do variable selection to understand which covariates and their lag are important for the explainability of water table depth. We could not find any parallel methodology that does handle suitably three major flexibilities: a) prediction under sparsity assumption within coefficients, b) considers a time series autoregression framework, and c) allows lags present in both dependent and independent variables, apart from ``BigVAR'' in the large interpretable family of linear models. We found the prediction for the daily level data is pretty well. In different practical scenarios, the monthly average prediction is not recommendable based on monthly storage data as it does not capture the daily level variability. We also obtained a similar inference in the table \ref{tab:monthly_prediction} where the adjusted R square is approximately poor with a minimum of .14 in comparison with the daily level and seasonal variation with a minimum adjusted square of .90. We recommend daily level prediction, the growing and dormant seasonal level prediction through our model. In the literature, for example, \cite{amatya2001hydrologic} explored water table height with the model DRAINMOD in a drained pine plantation of poorly drained soil. The model was based on drainage ditches, simulates interception, and evapotranspiration (ET) as the sum of canopy transpiration and soil evaporation, drainage, and surface runoff with the data from 1988 to 1997. As water table height is a highly correlated variable with water table depth, we may consider the comparison with the results from \cite{amatya2001hydrologic} assuming similar performance in our prediction interval. The crucial observation was that the $R^2$ varies from .65 to .91 with the majority of years 60-70\% explainability across different years. In comparison to that, our performance was always more than 90\% for daily, growing, and dormant seasons after considering a time series model. In \cite{latimer2022enhancing} page 177, it was found that another well discussed model, MIKE SHE only obtained NSE as .45 for daily water table depth prediction whereas using a statistical model ``BigVAR'', we obtained more than 90\% NSE for daily, growing and dormant season. So in comparison to the available methods, such as MIKE SHE and DRAINMOD, auto regressive process based prediction is stable consistently and performing better over a large time interval in the recent testing period across four different spatial locations for daily, growing and dormant season. In ``BigVAR'' we found that previous couple of available water table depth can be used to predict future water table depth significantly through auto regressive process. So it depends only fewer external parameters and not limited to upland or lowland predictions described in \cite{amatya2024hydrometeorological}.

In the time series plots, for example, for WS80 in the years around 2004 and 2006 (figure \ref{fig:ts_plots_80_NC} ), for WS78 in the year 2018, and for WS77 in the year 2017 (figure \ref{fig:ts_plots_77_78}), we may observe that in few places $\Delta$ and $\times$ are away from each other. The reason is that when there is too much rainfall, the water table rises up, sometimes above the ground level 0 (>0). As more water is added to the saturation zone, the water table moves closer to the ground surface. When the water table depth is close or above the surface ($\geq0$), the water table's phenomena behave differently. So we do focus majorly in the water table data where the water table depth is below the surface (<0). Similar observations and explorations were already concluded in \cite{amatya2000effects} and \cite{amatya2024hydrometeorological} based on different types of soils and extreme weather conditions. In the daily level time series plots we can observe that the confidence interval in D1 (NC) is thicker than the daily time series prediction in WS77, WS78, and WS80. It helps us to interpret that there is more variability in D1 (NC) than in the three stations in SC. We may also observe the heteroscedasticity in D1 compared to watersheds in SC in figure \ref{fig:regression_plot}. Similar observations were also found in figure 8 of \cite{amatya2020long} that daily water table depth does significantly vary across WS80 in SC and D1 in NC form 2003 to 2008.\\
The conclusions are drawn as follows:
\begin{itemize}
\item BigVAR is a tools for modeling sparse high-dimensional multivariate time series. AR, VAR, and VARX are technical approaches that can help predict water table depth. 
\item Bigvar can be utilized as a predictive model for water table depth prediction with reasonably good performance and interpretability of the variable selection.
\item Different climate variables can describe water table dynamics. The lag structure in the model plays a key role in estimating water table depth. Using the nearest previous water table depth estimates at any given time point can create a suitable predictive model for water table dynamics. Additionally, incorporating important weather variables with their lag structure, such as rainfall, solar radiation, net radiation, and wind direction, enhances the predictive performance.
\end{itemize}
\section{Recommendations and future goals}
\label{Recommendations and future goals}
We do highly recommend to utilize a time series auto regressive model where we can incorporate water table depth lag values in addition to different climate variables with different lag structures for predicting water table depth. This model is well developed in theory and practice. In many scenarios, practitioners include different covariates based on their experience. ``BigVAR'' allows a penalized regression to do variable selection for inference and utilize that model for prediction.
We want to explore the following directions in the future:
\begin{itemize}
\item Determine the instances and frequency within both the observed and simulated data where the water table depth remained within 30 cm for 14 consecutive days.
\item Explore the implementation of a hierarchical structure to manage missing data. In the current study, days with any missing variables were discarded. Investigate whether new data imputation methods could enhance our predictions.
\item Consider developing a more complex spatiotemporal model using a larger dataset from various locations to improve prediction and interpretation. Is it feasible to create a model for a broader region, such as the entire state of South Carolina or the United States? Incorporating factors such as the distance and elevation of oceans, different watersheds, and the locations of wells may lead to better predictive models.
\item Through the use of priors, Bayesian models can naturally introduce regularization. For instance, priors like the spike-and-slab or horseshoe priors encourage sparsity, effectively shrinking less important parameters towards zero without manually tuning penalties as in ridge or Lasso regressions. A new model development in parallel to ``BigVAR'' is also under consideration for future research.

\end{itemize}
\section{Acknowledgement}
\label{acknowledgement}
This research was partly supported by an appointment with the National Science Foundation (NSF) Mathematical Sciences Graduate Internship (MSGI) Program. This program is administered by the Oak Ridge Institute for Science and Education (ORISE) through an interagency agreement between the U.S. Department of Energy (DOE) and NSF. ORISE is managed for DOE by ORAU. All opinions expressed in this paper are the author's and do not necessarily reflect the policies and views of NSF, ORAU/ORISE, or DOE. The authors also acknowledge Andy Harrison, Hydrology Technician, at US Forest Service Santee Experimental Forest for providing related data for South Carolina sites.
Similarly, thanks are due to the Forest Service Forest Watershed Research unit for providing on-site support to visit the monitoring sites at the Santee Experimental Forest watersheds. Alokesh Manna would also like to thank his advisor \href{https://en.wikipedia.org/wiki/Dipak_K._Dey}{Dr. Dipak Dey}, a Board of Trustees Distinguished Professor in the Department of Statistics at the University of Connecticut for several necessary guidelines for this project and recommendations for the MSGI program.

\bibliographystyle{apalike}
\bibliography{lgb1}
\end{document}